\documentclass[pre]{revtex4}

\begin{document}

\title{Synchronization Landscapes in Small-World-Connected Computer Networks}

\author{H. Guclu\footnote{Permanent address:
Center for Nonlinear Studies, Theoretical Division,
Los Alamos National Laboratory, MS-B258, Los Alamos, NM, USA, 87545}}
\author{G. Korniss}
\affiliation{Department of Physics, Applied Physics, and Astronomy,
Rensselaer Polytechnic Institute, 110 8$^{th}$ Street, Troy, NY, USA, 12180-3590}

\author{M. A. Novotny}
\affiliation{Department of Physics and Astronomy, and
Center for Computational Sciences,
Mississippi State University, P.O. Box 5167,
Mississippi State, MS, USA, 39762-5167}

\author{Z. Toroczkai}
\affiliation{Center for Nonlinear Studies, Theoretical Division,
Los Alamos National Laboratory, MS-B258, Los Alamos, NM, USA, 87545}

\author{Z. R\'acz}
\affiliation{Institute for Theoretical Physics - HAS,
E\"otv\"os University, P\'azm\'any s\'et\'any 1/a, 1117 Budapest, Hungary}


\begin{abstract}

Motivated by a synchronization problem in distributed computing we
studied a simple growth model on regular and small-world networks,
embedded in one and two-dimensions. We find that the synchronization
landscape (corresponding to the progress of the individual
processors) exhibits Kardar-Parisi-Zhang-like kinetic roughening
on regular networks with short-range communication links. Although
the processors, on average, progress at a nonzero rate, their
spread (the width of the synchronization landscape) diverges with
the number of nodes (desynchronized state)  hindering efficient
data management. When random communication links are added on top
of the one and two-dimensional regular networks (resulting in a
small-world network), large fluctuations in the synchronization
landscape are suppressed and the width approaches a finite value
in the large system-size limit (synchronized state). In the
resulting synchronization scheme, the processors make
close-to-uniform progress with a nonzero rate without global
intervention. We obtain our results  by ``simulating the
simulations", based on the exact algorithmic rules, supported by
coarse-grained arguments.


\end{abstract}


\date{\today}

\maketitle


\begin{section}{Introduction}

The study of complex networks pervades various areas of science ranging from sociology
to statistical physics \cite{ALBERT02,DOROGOVTSEV02,NEWMAN03}. A network in terms of modeling
can be defined as a set of
items, referred to as nodes with links connecting them.
Examples of real life complex networks include the Internet \cite{FALOUTSOS99,BARABASI99},
the World Wide Web \cite{ALBERT99}, metabolic networks \cite{JEONG00},
transportation networks \cite{VESPIG2004,AMARAL2005}, and
social networks \cite{EUBANK04}.

Regular lattices are commonly used to study physical systems
with short-range or long-range interactions.
Earlier network studies focused mostly on the topological
properties of the networks. Recent works, motivated by a large number of
natural and artificial systems, such as the ones listed above, have turned the focus
onto processes on networks, where the interaction and dynamics between the
nodes are facilitated by a complex network. The question then
is how the collective behavior of the system is influenced by
this possibly complex interaction topology. Watts and Strogatz,
inspired by a sociological experiment \cite{MILGRAM67}, have
proposed a network model known as the small-world (SW)
network \cite{WATTS98}. The SW concept describes
the observation that, despite their often large size, there is a relatively
short path between any two nodes in most networks with some degree of randomness.
The SW model was originally constructed as a network to
interpolate between regular lattices and completely random
networks \cite{ERDOS60}. Systems and models (with well known behaviors on regular
lattices) have been studied on SW networks, such as the Ising model
\cite{SCALETT91,GITTERMAN00,BARRAT00,Wheeler}, the XY model \cite{KIM01},
phase ordering \cite{HONG02_2}, the Edwards-Wilkinson model
\cite{KOZMA03,KOZMA05,KOZMA05b}, diffusion
\cite{MONASSON99,BLUMEN_2000a,BLUMEN_2000b,ALMAAS_2002,KOZMA03,KOZMA05,KOZMA05b,HASTINGS04},
and resistor networks \cite{KORN_PLA_swrn}.
Closely related to phase transitions and collective phenomena is
synchronization in coupled multi-component systems \cite{WIESENFELD96}.
SW networks have been shown to
facilitate autonomous synchronization which is an important feature of these networks
from both fundamental and system-design points of view \cite{STROGATZ01,BARAHONA02,HONG02_1}.
In this paper we study a synchronization problem which emerges \cite{KORN00_PRL} in
certain parallel distributed algorithms referred to as  parallel
discrete-event simulation (PDES) \cite{FUJIMOTO90,NICOL94,LUBA00,KOLA_REV_2005}.
First, we find that constructing a SW-like synchronization network for PDES can
have a strong impact on the scalability of the algorithm \cite{KORN03_SCI}.
Secondly, since the particular problem is effectively ``local" relaxation in a noisy environment
in a SW network, our study also contributes to the understanding of collective phenomena on
these networks.

Simulation of large spatially extended complex systems in physics, engineering,
computer science, or military applications require vast amount of
CPU-time on serial machines using sequential algorithms. PDES
enabled researchers to implement faithful simulations on
parallel/distributed computer systems, namely, systems composed of
multiple interconnected computers
\cite{FUJIMOTO90,NICOL94,LUBA00,KOLA_REV_2005}. PDES is a subclass
of parallel and distributed simulations in which changes in the
components of the system occur instantaneously from one state to
another. These changes are referred to as discrete events, e.g.,
spin-flip attempts in magnetic Ising models with Glauber dynamics.
The primary motivation in PDES is to perform parallel simulation
for large systems without altering the original physical dynamics.
In the physics, chemistry and biology communities these types of
simulations are most commonly referred to as dynamic or kinetic
Monte Carlo simulations. Examples for real-life complex systems
where discrete-event models are applicable
include cellular communication networks \cite{LUBA00,GREEN94},
magnetic systems \cite{KORN99_JCP,KORN01_PRE}, spatial epidemic
models \cite{EUBANK04,DEEL96}, thin-film growth
\cite{AMAR_PRB_2005a,AMAR_PRB_2005b,AMAR_JCP_2005}, battle-field models
\cite{NICOL88}, and internet traffic models \cite{COWIE99}. In
the above examples the discrete events are call arrivals, spin-flip
attempts, infections, monomer depositions, troop movements, and packet
transmissions/receptions respectively.

PDES has two basic ingredients: the set of local simulated times
of the processors or processing elements (PE) (also referred to as
\textit{virtual times} \cite{JEFF85}) and a synchronization scheme
\cite{FUJIMOTO90}. The difficulty in PDES is that the discrete
events are not synchronized by a global clock since the dynamic
is usually asynchronous. The challenge is algorithmically
parallelizing the physically non-parallel dynamics, while enforcing
causality between events and reproducibility. There are two main
approaches in PDES: ({\it i}) \textit{conservative
synchronization}, which avoids the possibility of any type of
causality errors by checking the causality relation between potentially related events at
every update attempt \cite{CHANDY79,CHANDY81} and ({\it ii})
\textit{optimistic synchronization}, which allows possible causality
errors, then later initiates {\it rollbacks} to correct the erroneous
computations \cite{JEFF85,DEELMAN97}. Innovative
methods have also been introduced to make optimistic
synchronization more efficient, such as reverse computation
\cite{CAROTHERS99}. Other recent improvements to exploit
parallelism in discrete-event systems are the ``lookback" method
\cite{CHEN02} and the freeze-and-shift algorithm \cite{SHCHUR04}.

As the number of available PEs on parallel architectures increases
to tens of thousands \cite{TOP500_IBM}, and high-performance
grid-computing networks emerge \cite{GRID,KIRKPATRICK03},
fundamental questions of the scalability of the underlying
algorithms must be addressed. A PDES should have the following
properties to be scalable \cite{GREEN94}: ({\it i}) the virtual
time horizon formed by the virtual times of the PEs should
progress on average with a non-zero rate and ({\it ii}) the
typical spread (width) of the time horizon should be bounded as
the number of PEs, $N_{PE}$, goes to infinity. The latter condition
becomes important to avoid long delays while waiting for ``slow"
nodes \cite{AMAR_PRB_2005a} or, alternatively, to eliminate the
need to reserve a large amount of memory for temporary data storage.
In this paper we study regular and SW network communication
topologies and show a possible way to construct \textit{fully}
scalable parallel algorithms for systems with
\textit{asynchronous} dynamics and short-range interactions on
regular lattices.

In order to understand scalability and synchronizability of PDES,
we consider the parallel simulation itself as a complex
interacting system where the specific synchronization rules
correspond to the ``microscopic dynamics''. A similar approach was
also successful to establish a connection \cite{SLOOT01} between
rollback-based (or optimistic) schemes \cite{JEFF85} and
self-organized criticality \cite{BAK87,BAK88}. Our approach
exploits a mapping between non-equilibrium surface growth and the
evolution of the simulated time horizon \cite{KORN00_PRL,KORN00_UGA}
so that we can use the tools and framework of statistical
mechanics. A similar analogy between phase transitions and
computational complexity has turned out to be highly fruitful to
gain more insight into traditional hard computational problems
\cite{COMPLEX,MONA99}. In this paper we consider the scalability
of conservative synchronization schemes for self-initiating
processes \cite{NICOL91,FELDER91}, where update attempts on each
node are modeled as independent Poisson streams. We study the
morphological properties of the simulated time horizon (or
synchronization landscape). Through our study one also gains some
insight into the effects of SW-like interaction topologies on the
critical fluctuations in interacting systems.

This paper, in part, is an expanded version of our earlier works on
one-dimensional (1D) \cite{KORN00_PRL} and 1D SW (SW embedded in
one dimension) synchronization networks \cite{KORN03_SCI}.
Further, we present new results for the two-dimensional (2D), and
for the 2D SW (SW embedded in two dimensions) synchronization
networks. The paper is organized as follows. In Section II we show
detailed results for a previously studied short-range model with
nearest neighbor communication which we refer to as the basic
conservative synchronization (BCS) scheme in one dimension
\cite{KORN00_PRL}. In Section III we present results on SW
networks constructed by adding random links on a regular
one-dimensional network \cite{KORN03_SCI}. Section IV and Section
V present the results on the BCS scheme in two dimensions and its
SW version, respectively. In Section VI we summarize our work and
discuss results.

\end{section}


\begin{section}{Basic Conservative Synchronization Scheme in 1D}

First, we briefly summarize the basic observables relevant to our
analysis for synchronization and their scaling relations borrowed
from non-equilibrium surface growth theory. The set of local
simulated times for the PEs, $\{\tau_i(t)\}_{i=1}^{N_{PE}}$,
constitutes the simulated time horizon. Here $N_{PE}$ is the
number of PEs and $t$ is the discrete number of parallel steps,
proportianal to the real/wall-clock time. On a regular
$d$-dimensional hypercubic lattice $N_{PE}$=$N^d$, where $N$ is
the linear size of the lattice. For a one-dimensional system
$N_{PE}$=$N$. For the rest of the paper we will use the term
``height'', ``simulated time'', or ``virtual time''
interchangeably, since we refer to the same local observable
(local field variable).

Since the discrete events in PDES are not synchronized by a global
clock, the processing elements have to synchronize themselves by
communicating with others. One of the first approaches to this
problem for self-initiating processes \cite{NICOL91,FELDER91} is
the BCS scheme proposed by Lubachevsky \cite{LUBA87,LUBA88} by
using only nearest neighbor interactions, mimicking the
interaction topology of the underlying physical system
\cite{KORN00_PRL}. His basic model associates each component or
site with one PE (worst-case scenario) under periodic boundary
conditions. In this BCS scheme, at each time step, only those PEs
whose local simulated time is {\em not larger} than the local
simulated times of their next nearest neighbors are incremented by
an exponentially distributed random amount so that the discrete
events exhibit Poisson asynchrony. Otherwise (if the local
simulated time of the PE is larger than any of its neighbors'
simulated time), no update occurs, i.e., the PE idles. The
evolution equation for the local simulated time of site $i$ simply
becomes
\begin{equation}
\tau_i(t+1) = \tau_i(t) + \eta_{i}(t)
\Theta\left(\tau_{i-1}(t)-\tau_i(t)\right)\Theta\left(\tau_{i+1}(t)-\tau_i(t)\right)\;,
\label{evolution}
\end{equation}
where $\eta_{i}(t)$ is an exponentially distributed random number and
$\Theta(...)$ is the Heaviside step-function. In one-dimension with
periodic boundary conditions, the network has a ring topology as
shown in Fig.~\ref{fig_1dmodel}(a), so each node is connected to
the nearest left and right neighbors. The nearest-neighbor
interaction in the BCS scheme implies that in order to ensure
causality, PEs need to exchange their local simulated (virtual)
times only with neighboring PEs in the virtual network topology.
The possible configurations for the local simulated times for the
successive nodes are shown in Fig.~\ref{fig_slopes}. In these
configurations an update occurs only if the node we are
considering (node $i$) is a local minima. In the other three cases
the node $i$ idles. The algorithm is obviously free from deadlock,
since at worst, the PE with the absolute minimum local simulated
time can make progress. Note that from an actual parallel
implementation viewpoint, equal virtual times (and hence
conflicting updates) are extremely unlikely (to the extent of the
resolution of generating continuously distributed exponential
variables), and the algorithm allows updates if the neighboring
virtual times are greater than or equal to that of PE $i$. From a
theoretical viewpoint, for $t$$>$$0$, the probability density of
the simulated time horizon $\{\tau_{i}(t)\}_{i=1}^{N_{PE}}$ is a
continuous measure, so the probability that the simulated times of
any two nodes are the same is of measure zero.

In analyzing the performance of the above scheme, it is
helpful that the progress of the simulation itself is decoupled from
the possibly complex behavior of the underlying system. This is contrary to
optimistic approaches, where the evolution of the underlying system and the progress
of the PDES simulation are strongly entangled \cite{SLOOT01}, making scalability
analysis a much more difficult task.
\begin{figure}[htb]
\vspace{5cm}
\includegraphics{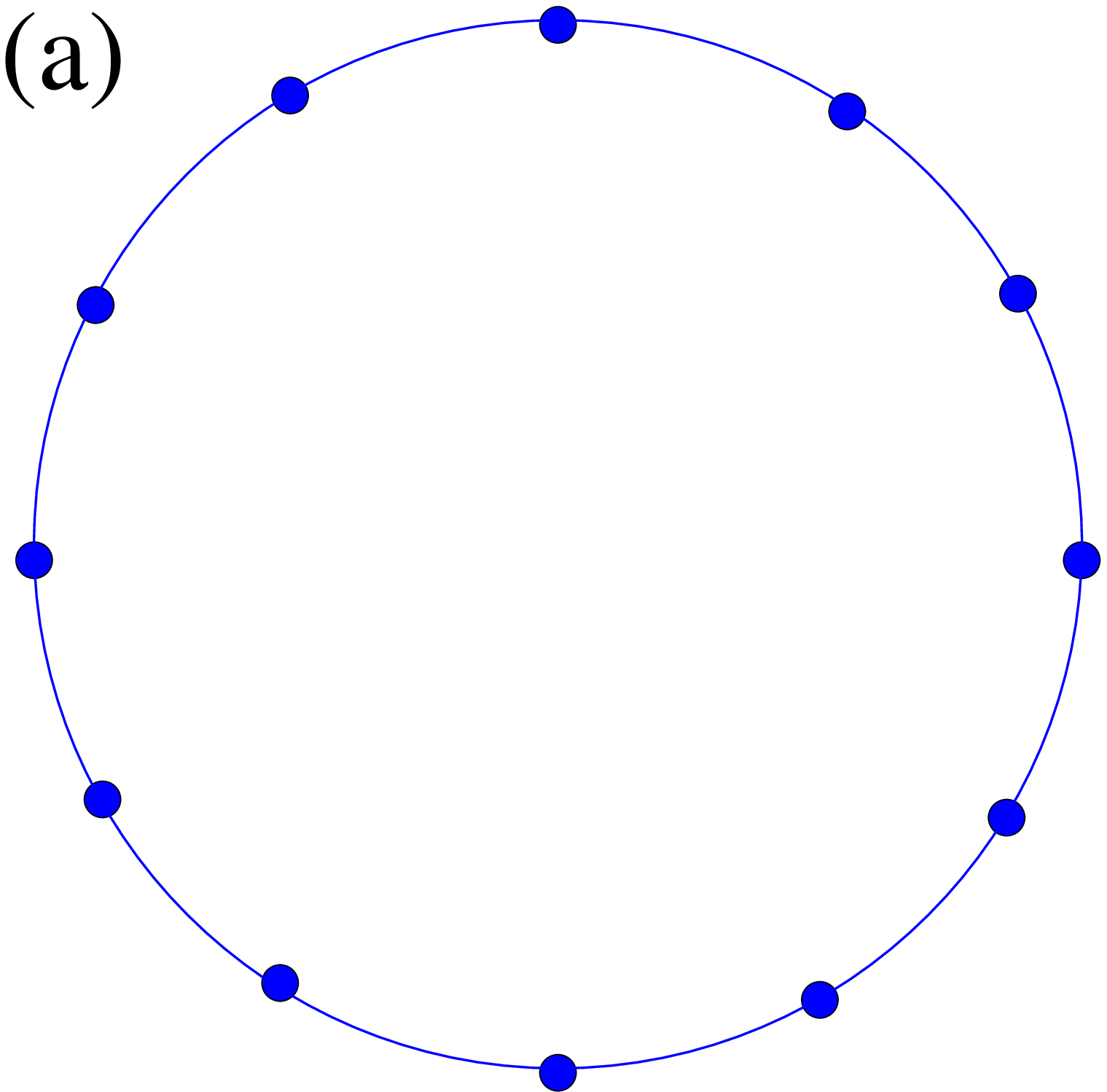}
\includegraphics{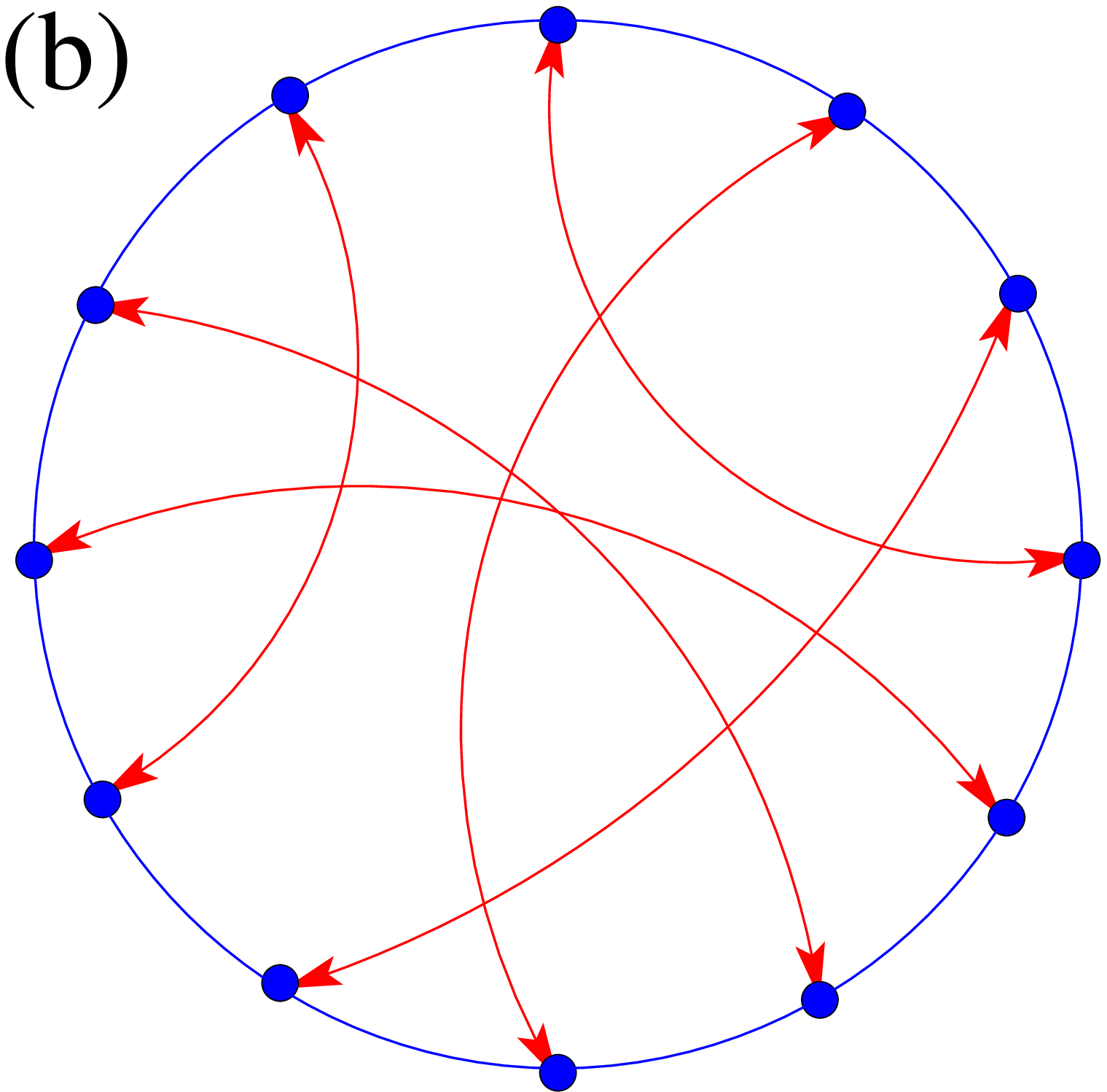}
\caption{(a) One-dimensional (1D) regular network (with periodic boundary conditions), where nodes
are connected to their nearest neighbors.
(b) Small-world (SW) synchronization network, where each node is connected to exactly one
randomly chosen node, in addition to the nearest neighbors.}
\label{fig_1dmodel}
\end{figure}

One of the important aspects of conservative PDES is the
theoretical efficiency or \textit{utilization} which is
defined as the average fraction of non-idling PEs at each
parallel step. This measure also provides the average progress
rate of the simulation, as can be seen from Eq.~(\ref{evolution}).
In the BCS scheme, where only nearest-neighbor interactions are
present, the utilization is equal to the density of local minima
in the simulated time horizon. Thus, the utilization (on a regular
one-dimensional lattice) can be written as
\begin{equation}
\langle u \rangle =
\langle\Theta(\tau_{i-1}-\tau_i)\Theta(\tau_{i+1}-\tau_i)\rangle =
\langle \Theta(-\phi_{i-1})\Theta(\phi_i)\rangle \;,
\label{utilization}
\end{equation}
where $\phi_i \equiv \tau_{i+1}-\tau_i$ is the local slope,
$\Theta(...)$ is the Heaviside step function, and $\left\langle
... \right\rangle$ denotes an ensemble average over the noise in
Eq.~(\ref{evolution}). The utilization is independent of $i$ for a
system of identical PEs due to the translational invariance.
\begin{figure}[htb]
\vspace{4cm}
\includegraphics{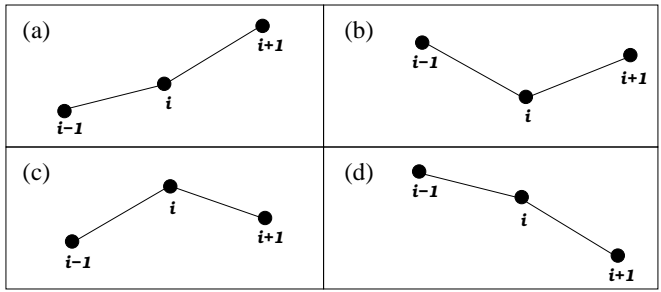}
\caption{Possible simulated-time configurations for three successive nodes
(involving two successive slopes) in the basic conservative scheme (BCS) in one dimension.
From the perspective of node $i$, only configuration (b) allows it to proceed
(node $i$ is a local minimum). In all other cases causality could be violated if
an update occurs at site $i$, because the local field variables of the neighboring nodes
are not known at the instant of update attempt.}
\label{fig_slopes}
\end{figure}

Another important observable of PDES is the statistical spread or \textit{width} of
the simulated time surface. The width
of the simulated time surface $w$ is defined as the root mean square of the virtual
times with respect to the mean, $w=\sqrt{\langle w^2 \rangle}$, where
\begin{equation}
\left\langle w^{2} \right\rangle \equiv \left\langle w^{2}(N,t) \right\rangle =
\left\langle \frac{1}{N^d}\sum_{i=1}^{N^d}
\left[ \tau_i(t)-\bar{\tau}(t) \right]^2 \right\rangle \;\;,
\label{width_def}
\end{equation}
and $\bar{\tau}(t)$$=$$(1/N^d)\sum_{i=1}^{N^d} \tau_i(t)$ is the mean progress
(``mean height") of the time surface.

Since we use the formalism and terminology of non-equilibrium surface growth phenomena,
we briefly review scaling concepts for self-affine or rough surfaces. The scaling behavior
of the width, $\langle w^{2}(N,t) \rangle$, alone typically captures and
identifies the universality class of the non-equilibrium growth process
\cite{BARA95,HEALY95,KRUG97}. In a finite system the width initially grows as
$\left\langle w^2(N,t)\right\rangle$$\sim$$t^{2\beta}$, and after a system-size
dependent cross-over time $t_{\times}$$\sim$${N}^{z}$, it reaches a steady-state
$\left\langle w^2(N,t)\right\rangle$$\sim$${N}^{2\alpha}$ for $t$$\gg$$t_x$. In the relations above
$\alpha$, $\beta$, and $z$=$\frac{\alpha}{\beta}$ are called the roughness,
the growth, and the dynamic exponent, respectively. The temporal and system-size
scaling of the width can be captured by the Family-Vicsek \cite{FAMILY85} relation,
\begin{equation}
\left\langle w^2(N,t)\right\rangle = N^{2\alpha}f(t/N^{z})
\;.
\label{vicsek}
\end{equation}
Note that the scaling function $f(x)$ depends on $t$ and the linear system-size $N$
only through the specific combination $x=t/N^z$, reflecting the importance of the
crossover time $t_\times$. For small values of its argument $f(x)$ behaves as a power
law, while for large arguments it approaches a constant,
\begin{equation}
f(x) \sim
\left\{
\begin{array}{ll}
x^{2\beta}    & \mbox{if $x$$\ll$$1$} \\
\mbox{const.} & \mbox{if $x$$\gg$$1$}
\end{array}
\right. \;,
\label{f_scaling}
\end{equation}
yielding the correct scaling behavior of the width for early times ($x$$\ll$$1$) and for the
steady-state ($x$$\gg$$1$).

A somewhat less frequently studied quantity is the growth rate of a
growing surface. This quantity is typically non-universal
\cite{KORN00_PRL,TORO00,KRUG90,KORN_ACM,KOLA03_PRE_1,KOLA03_PRE_2,KOLA04},
but as was shown by Krug and Meakin \cite{KRUG90}, on
$d$-dimensional regular lattices, the finite-size corrections to it
are. In the context of the basic PDES scheme, the growth rate of the
simulated time surface corresponds to the progress rate (or
utilization) of the simulation, hence our special interest in this observable.
For the finite-size behavior of the steady-state growth rate, one has \cite{KRUG90}
\begin{equation}
\langle u(N)\rangle \simeq \langle u(\infty)\rangle + \frac{const.}{N^{2(1-\alpha)}} \;,
\label{utild}
\end{equation}
where $\langle u(\infty) \rangle$ is the value of the growth rate
in the asymptotic infinite system-size limit and
$\alpha$ is the dimension-dependent roughness exponent of the growth process.

\begin{figure}[htb]
\vspace{7cm}
\includegraphics{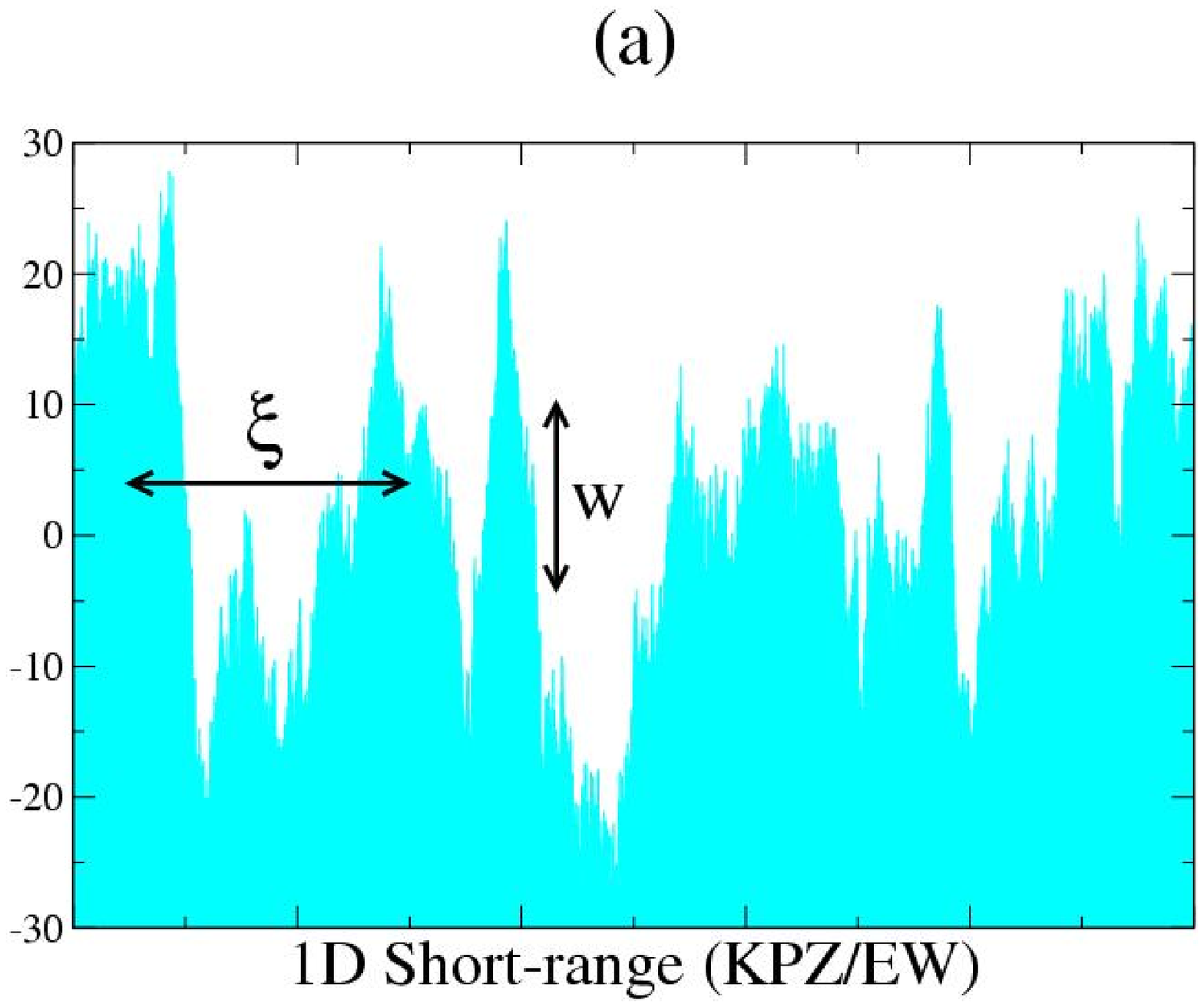}
\includegraphics{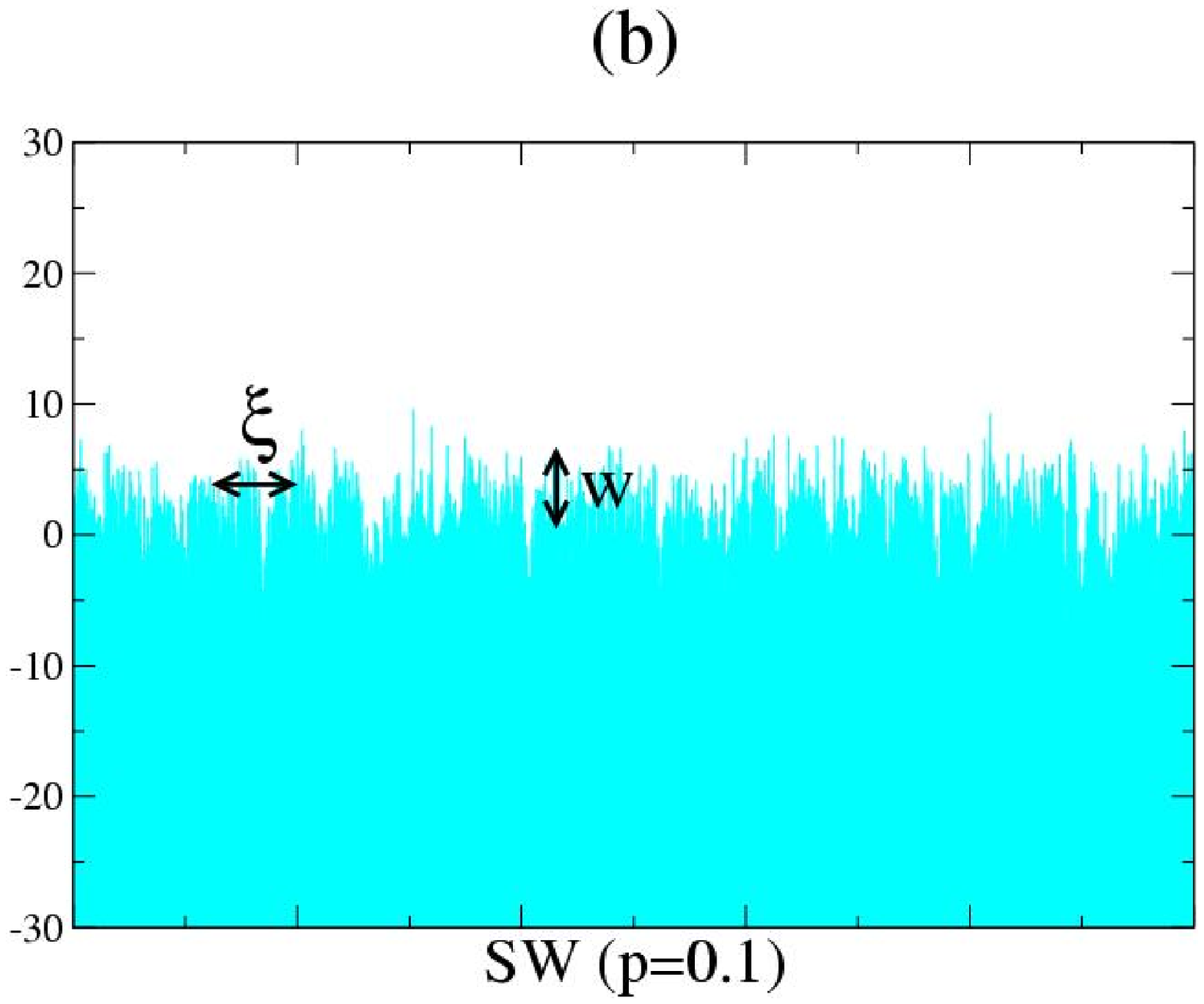}
\includegraphics{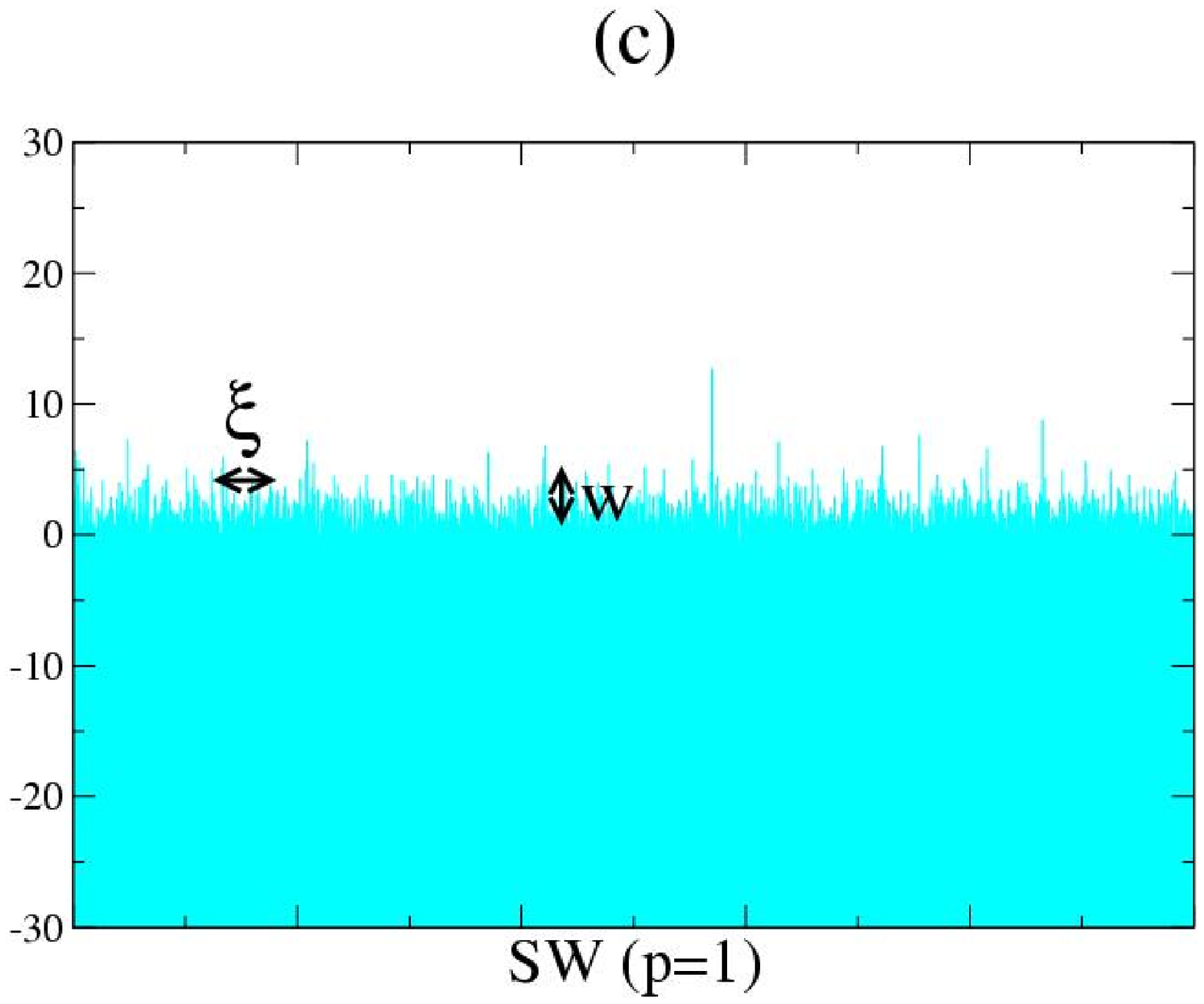}
\vspace{-2cm}
\caption{Virtual time horizon snapshots for 10,000 sites in 1D.
(a) For the regular network with nearest-neighbor connections ($p$=$0$).
The lateral correlation length $\xi$ and width $w\equiv \sqrt{w^2}$ are
shown for illustration in the graph. The rough steady-state surface belongs to the
KPZ/EW universality class.
For the SW synchronization network with (b) $p$$=$$0.1$ and (c) $p$$=$$1$
the heights are effectively decorrelated and both
the correlation length and the width are reduced, and approach system-size independent values
for sufficiently large systems. The resulting surface is macroscopically smooth.
Note that the heights are relative to the average height.}
\label{figsn}
\end{figure}

Based on a mapping between virtual times and surface heights
\cite{KORN00_PRL}, in the coarse-grained description, the virtual
time horizon of the BCS was found to be governed
\cite{KORN00_PRL,TORO00} by the Kardar-Parisi-Zhang (KPZ) equation
\cite{KARD86}, well-known in surface growth phenomena
\begin{equation}
\partial_t \hat{\tau}_i = \nabla^2\hat{\tau}_i - \lambda (\nabla \hat{\tau}_i)^2
+ ... + \eta_i(t)
\;,
\label{KPZ}
\end{equation}
where $\nabla^2$ stands for the discretized Laplacian,
$\nabla^2\hat{\tau}_i = \hat{\tau}_{i+1} + \hat{\tau}_{i-1} - 2\hat{\tau}_i$,
$\nabla$ is the discretized gradient operator,
$\nabla \hat{\tau}_i = \hat{\tau}_{i+1}-\hat{\tau}_i$,
$\hat{\tau}_i=\tau_i-\bar{\tau}$ is the surface height fluctuation
(or virtual time) measured
from the mean, $\eta_i(t)$ is a noise delta-correlated in space and time,
$\langle\eta_i(t)\eta_j(t')\rangle = 2D\delta_{ij}\delta(t-t')$,
$\lambda$ is a positive constant and $...$ stands for higher order irrelevant terms
(in a renormalization group sense).
Equation~(\ref{KPZ}) can also give an account of a number
of other nonlinear phenomena such as Burgers turbulence and directed polymers in
random media \cite{BARA95}. When the simple update rule of the
basic synchronization scheme is implemented on a one-dimensional regular network,
one can observe a simulated time surface governed by the KPZ equation,
and in the steady-state,
by the Edwards-Wilkinson Hamiltonian \cite{EDWARDS82} [Fig.~\ref{figsn}(a)].

When analyzing the statistical and morphological properties of the
stochastic landscape of the simulated times, it is convenient to
study the height-height correlation or its Fourier transform, the
height-height structure factor. The equal-time height-height
structure factor $S(k,t)$ in one-dimension is defined as
\begin{equation}
S(k,t)N\delta_{k,-k^\prime} = \langle \tilde{\tau}_k(t)\tilde{\tau}_{k^\prime}(t)
\rangle
\;,
\label{str_factor}
\end{equation}
where $\tilde{\tau}_k = \sum_{j=1}^{N}e^{-ikj}\tau_j$ is the
Fourier transform of the virtual times with the wave number
$k$=$2\pi n/N$, $n$=$0,1,2,...,N-1$ and $\delta_{k,-k^\prime}$ is
the Kronecker delta. The structure factor essentially contains all the
``physics'' needed to describe the scaling behavior of the time
surface. Here we focus on the steady-state properties
($t$$\to$$\infty$) of the time horizon where the structure factor
becomes independent of time, $\lim_{t\to\infty}S(k,t)=S(k) $. In
the long-time limit in one dimension, for a KPZ surface described
by Eq.~(\ref{KPZ}) one has \cite{TORO00}
\begin{equation}
S(k) = \frac{D}{2[1-\cos(k)]} \sim \frac{1}{k^2}\;,
\label{sf_toro}
\end{equation}
where the latter approximation holds for small values of $k$. By
performing the inverse Fourier transformation of
Eq.~(\ref{sf_toro}), we can also obtain the spatial two-point
function,
\begin{equation}
G(l) = (1/N)\sum_{k\neq 0}e^{ikl}S(k)\;,
\label{G_l}
\end{equation}
yielding \cite{TORO00,TORO03}
\begin{equation}
G(l)\simeq\frac{D}{2}\left(\frac{N}{6} - l \right)
\label{corr_toro}
\end{equation}
for $1\ll l \ll N$. In particular, for the steady-state width one finds
\begin{equation}
\langle w^2 \rangle = \frac{1}{N}\sum_{k \not= 0} S(k) =G(0)
\simeq \frac{D}{12}N \sim N
\label{sf_w}
\end{equation}
in one dimension \cite{TORO00}. This divergent width is caused by
a divergent length scale, $\xi$, the ``lateral" correlation length
in the KPZ-like synchronization landscape.

The measured steady-state structure factor
[Fig.~\ref{fig_kpz1dsf}(a)], obtained by simulating the BCS based
on the exact rules for the evolution of the synchronization
landscape, confirms the coarse-grained prediction for small $k$
values, $S(k) \sim 1/k^2$. Figure~\ref{fig_kpz1dsf}(b) shows the
corresponding spatial two-point correlation function $G(l)$.
\begin{figure}[htb]
\vspace{6cm}
\includegraphics{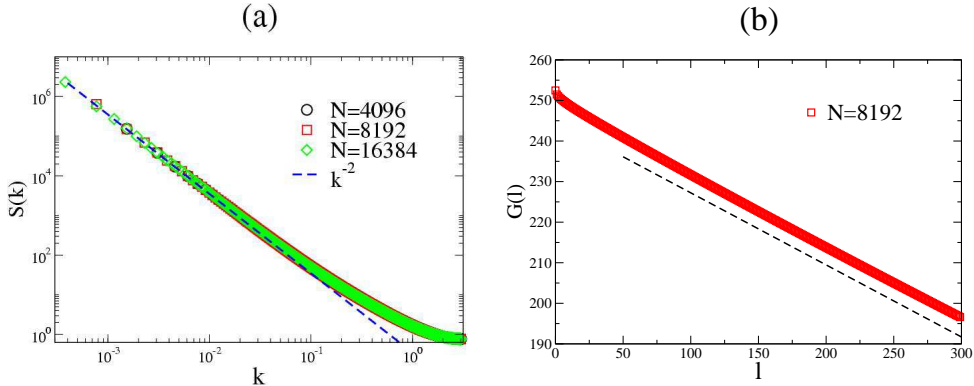}
\includegraphics{1D-KPZ-corr-L8192.eps}
\caption{(a) The steady-state structure factor as a function of the wave number for the BCS
scheme in 1D. The small-$k$ course-grained prediction (consistent with the steady-state EW/KPZ
universality class in 1D) is indicated by a dashed line [Eq.~(\ref{sf_toro})].
Note the log-log scales.
(b) Steady-state spatial two-point correlation function. The straight line again indicates
the asymptotic EW/KPZ behavior in one dimension [Eq.~(\ref{corr_toro})].}
\label{fig_kpz1dsf}
\end{figure}
Simulations of the BCS scheme in one dimension yield scaling
exponents that agree within error with the predictions of the KPZ
equation \cite{BARA95,HEALY95,KARD86}. The time evolution of the width
[Fig.~\ref{fig_kpz1dw2Lt}(a)] shows that the growth exponent
$\beta$$\simeq$$1/3$.  Looking at the system-size dependence of the
steady-state width [Figure~\ref{fig_kpz1dw2Lt}(b)], we find the
roughness exponent $\alpha$$\simeq$$1/2$, consistent with the
one-dimensional KPZ value, $\langle w^2 \rangle$$\sim$$N^{2\alpha}$$\sim$$N$.
The dynamic exponent values found from the width as a
function of the cross-over time and $z$=$\alpha/\beta$ are the same,
about $3/2$. The inset in Fig.~\ref{fig_kpz1dw2Lt}(a) shows
that the scaled version of the width evolution by using the
scaling exponents is consistent with the Family-Vicsek relation
[Eq.~(\ref{vicsek})], although with relatively large corrections
to scaling.

The width distributions, $P(w^2)$, have been introduced to provide
a more detailed characterization of surface growth processes
\cite{FOLTIN94,PLISCHKE94,RACZ94,ANTAL96,ANTAL2001,ANTAL2002} and have been used
to identify universality classes \cite{KORN00_PRL}. The width
distribution of rough surfaces belonging to the same universality class
is governed by a universal scaling function $\Phi(x)$,
such that $P(w^2)$$=$$\langle w^2\rangle^{-1}\Phi(w^2/\langle
w^2\rangle)$. $\Phi(x)$ can be calculated analytically for a
number of models, including the EW class \cite{FOLTIN94}. The width
distribution for the basic synchronization scheme is shown in
Fig.~\ref{fig_kpz1dw2Lt}(c). Systems with $N$$\geq$$10^3$ show
convincing data collapse onto this exact scaling function. The
inset in Fig.~\ref{fig_kpz1dw2Lt}(c) shows the same graph in
log-normal scale to show the collapse at the tail of the
distribution. The convergence to the limit distribution is very
slow when compared to other microscopic models (such as the
single-step model \cite{BARA95,ANTAL96}) belonging to the same
universality class.
\begin{figure}[htb]
\vspace{5cm}
\includegraphics{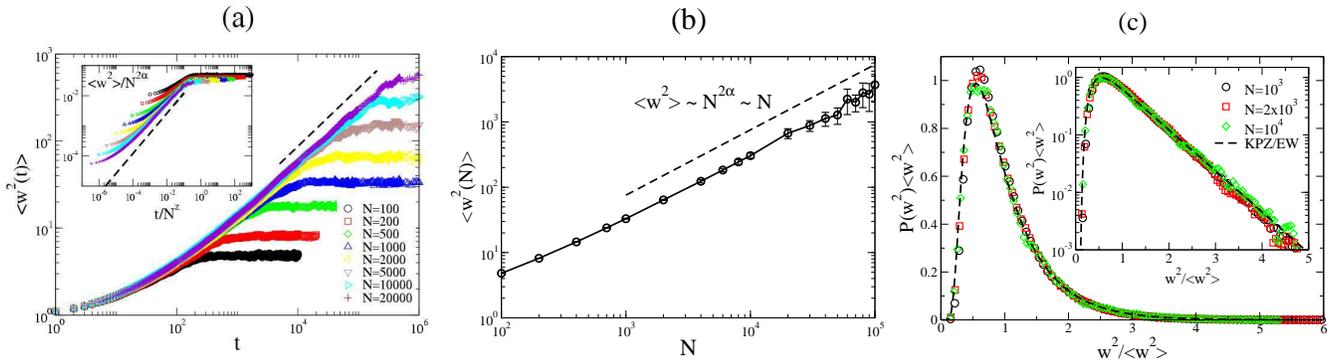}
\includegraphics{1D-KPZ-w2L.eps}
\includegraphics{1D-KPZ-w2-dist.eps}
\caption{(a) Time evolution of the width for different system-sizes
in the BCS scheme in 1D. The inset shows the same data on scaled axes,
$\langle w^2\rangle/N^{2\alpha}$ versus $t/N^z$.
Each curve has been obtained by averaging over at least fifty realizations.
(b) The steady-state width
of the time horizon for the one-dimensional BCS as a function
of system-size. The dashed straight line represents the asymptotic one-dimensional
KPZ/EW behavior, $\left\langle w^2(N)\right\rangle$$\sim$$N^{2\alpha}$ with $\alpha$$=$$1/2$.
(c) Scaled width distributions for the BCS scheme in 1D.
The exact asymptotic EW/KPZ width distribution \cite{FOLTIN94} is shown with a dashed line.
The inset shows the same distributions on log-normal scales.}
\label{fig_kpz1dw2Lt}
\end{figure}

Now we discuss our findings for the steady-state utilization of
the BCS scheme. As stated above, the synchronization landscape of
the virtual times belongs to the EW universality class in one
dimension. This implies that the {\em local slopes} in the
steady-state landscape are short-range correlated \cite{TORO00}.
Hence the density of local minima in the synchronization
landscape, and in turn the utilization, remains {\em nonzero} in
the infinite system-size limit \cite{KORN00_PRL,TORO00}. For
fixed $N$, the utilization drops from relatively higher initial
value at early times to its steady-state value in a very short
time [Fig.~\ref{fig_kpz1dutil}]. Further, the steady-state
utilizations for various systems converge to the asymptotic
system-size independent value.
\begin{figure}[htbp]
\vspace{6cm}
\includegraphics{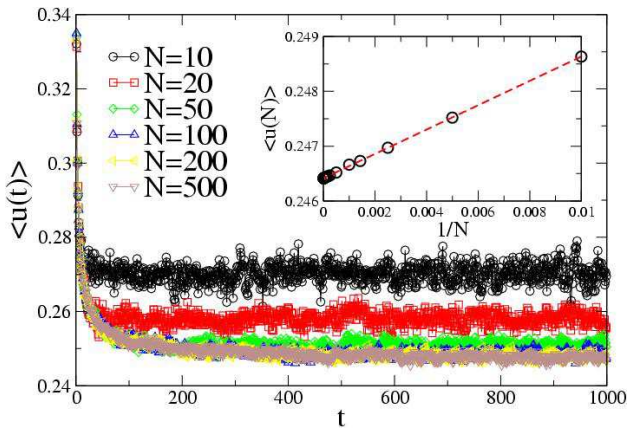}
\vspace{-0.3cm}
\caption{Utilization
in the 1D BCS scheme as a function of time for various system
sizes. The inset shows the steady-state utilization as a function
of the $1/N^{2(1-\alpha)}$ with the 1D KPZ roughness exponent
$\alpha$$=$$1/2$. The dashed line is a linear fit to
Eq.~(\ref{utild}) with $\alpha$$=$$1/2$,
$\langle u(N)\rangle \approx 0.2464 + 0.2219/N$.}
\label{fig_kpz1dutil}
\end{figure}
In 1D, where $\alpha$$=$$1/2$, the system-size dependence of the steady-state utilization
must follow [Eq.~(\ref{utild})]
$\langle u(N)\rangle \simeq \langle u(\infty)\rangle + \frac{const.}{N}$,
as confirmed by our direct simulations for the BCS scheme [inset
of Fig.~\ref{fig_kpz1dutil}]. For the KPZ model [Eq.~(\ref{KPZ})]
$\langle u(\infty)\rangle$$=$$1/4$, since in the steady state, the
local slopes are delta-correlated, resulting in a probability of
$1/4$ for the configuration in Fig.~\ref{fig_slopes}(b),
corresponding to a local minimum. For the actual BCS
synchronization profile $\langle u(\infty)\rangle$$\simeq$$0.2464$
\cite{KORN00_PRL,TORO00}, as a result of non-universal short-range
correlations present for the slopes in the specific microscopic
model \cite{TORO03}. Finally one can argue that the utilization
for the BCS on regular lattices remains \textit{nonzero}  in the
thermodynamic limit and it displays universal finite-size effects
\cite{KORN_ACM,KOLA03_PRE_1,KOLA03_PRE_2,KOLA04,KRUG90}. Note that
a small change with respect to the ``worst case" (one-site-per PE)
scenario, in particular, hosting more than one site per PE, leads
to utilizations that are close to the limiting value of unity
\cite{KOLA03_PRE_1,KOLA04}, and hence are practical for actual
PDES implementations
\cite{KORN99_JCP,AMAR_PRB_2005a,AMAR_PRB_2005b}.

In order to obtain an analytically tractable scalability model for
the BCS scheme, Greenberg et al. introduced the $K$-random
interaction network model \cite{GREEN96}. In this model at each
update attempt PEs compare their local simulated times with those
of a set of $K$ \textit{randomly} chosen PEs. This set is
re-chosen for each update attempt (i.e., the network is
``annealed"), even if a previous update attempt has failed. It was
shown that in the limit of $t$$\to$$\infty$ and $N$$\to$$\infty$,
the utilization (or the average rate of progress) converges to a
\textit{non-zero} constant, $1/(K+1)$. They also suggested that
the scaling properties of $K$-random model as $t$$\to$$\infty$ and
$N$$\to$$\infty$ are universal and hold for regular lattices as
well. But changing the interaction from nearest neighbor PEs on a
regular lattice to randomly chosen PEs changes the universality
class of the time horizon. Simply put, the underlying topology has
crucial effects on the universal behavior of the time horizon. The
random (annealed) interaction topology of the $K$-random model
results in a mean-field-like behavior, where the simulated time
surface is uncorrelated and has a {\em finite} width in the limit
of an infinite number of PEs. Their conjecture for the width does
not hold, thus, the BCS scheme for \textit{regular lattices}
cannot be equivalently described by the $K$-random model (at least
not below the upper critical dimension of the KPZ universality
class \cite{MARINARI02}). However, we were inspired by
\cite{GREEN96} to change the communication topology of the PEs by
introducing random links \textit{in addition} to the necessary
short-range connections. In the next section we present our
modification to the original conservative scheme on regular
lattices to achieve a fully scalable algorithm where both
scalability conditions are satisfied: the simulation has a nonzero
progress rate and the width of the synchronization landscape is
finite.

\end{section}


\begin{section}{Small-World Synchronization Scheme in 1D}

The divergent width for the larger systems, discussed in the previous section,
is the result of the divergent lateral correlation length $\xi$ of the virtual
time surface, reaching the system-size $N$ in the steady-state
\cite{BARA95,KORN_ACM,KOLA03_PRE_1,KOLA03_PRE_2,KOLA04}. To de-correlate the
simulated time horizon, first, we modify the virtual communication topology of the
PEs. The resulting communication network must include the original short-range
(nearest-neighbor) connections to faithfully simulate the dynamics of the
underlying system. In the modified network, the connectivity of the nodes
(the number of neighbors) should remain non-extensive (i.e., only a finite number
of virtual neighbors per node is allowed). This is in accordance with our desire
to design a PDES scheme where no global intervention or synchronization is employed
(PEs can only have $O(1)$ communication exchanges per step). It is clear that the
added synchronization links (or at least some of them) have to be long range.
Short range links alone would not change the universality class and the scaling
properties of the width of the time horizon. One can satisfy this condition
by selecting the additional links called ``small-world" links randomly among all
the nodes in the network. Also, fluctuations in the individual
connectivity should be avoided for load balancing purposes, i.e., requiring
the same number of added links (e.g., one) for each node is a reasonable constraint.
We then choose the extra synchronization links in such a way that each PE is connected
to exactly one other PE via a ``quenched" bidirectional link [Fig.~\ref{fig_1dmodel}(b)].

One of the basic structural characteristics of SW-like networks is the ``low degree
of separation" between the nodes. The most commonly used observables to analyze this
property are the average shortest path length, $\delta_{avg}(N)$, and the maximum
shortest path length, $\delta_{max}(N)$. The shortest path length between two nodes is
defined as the minimum number of nodes needed to visit in order to go from one of
the nodes to the other.
The average shortest path length is obtained by averaging the above quantity
between all possible pairs of nodes in a given network.
The maximum shortest path length, also known as the \textit{diameter} of the network,
is the length of the longest among the shortest paths in the network.
Both of these observables scale logarithmically
with the system-size $N$ in SW-like networks \cite{BOLLOBAS}. The system-size
dependence of these path lengths for our one-dimensional SW network is
logarithmic as expected, see Fig.~\ref{fig_1d-qrm-sp}.
\begin{figure}[htb]
\vspace{6.2cm}
\includegraphics{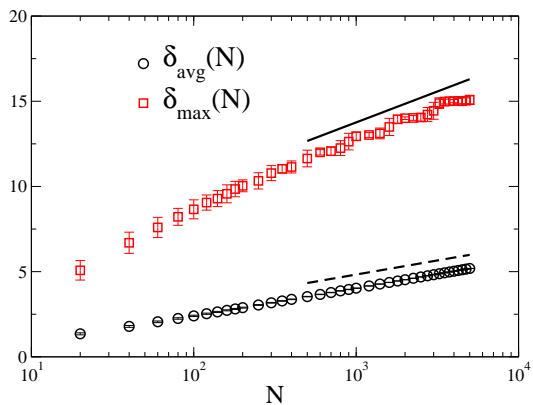}
\caption{Average and maximum shortest path (diameter) as a function of
the number of nodes for the SW synchronization network in 1D, as described in the text.
The latter is also referred to as the diameter of the network. The solid and dashed lines both
indicate the logarithmic dependence. Note the normal-log scales.}
\label{fig_1d-qrm-sp}
\end{figure}

We now describe the modified algorithmic steps for the SW
connected PEs \cite{KORN03_SCI}. In the SW conservative PDES
scheme, in every parallel time step, each PE, with probability $p$,
compares its local simulated time with its \textit{full} virtual
neighborhood, and can only advance if it is the minimum in this
neighborhood, i.e., if $\tau_i(t)$$\leq$$\min\{\tau_{i-1}(t),\tau_{i+1}(t),\tau_{r(i)}(t)\}$,
where $r(i)$
is the random connection of PE $i$. With probability $(1$$-$$p)$ each
PE follows the original scheme, i.e., the PE then can advance if
$\tau_i(t)$$\leq$$\min\{\tau_{i-1}(t),\tau_{i+1}(t)\}$. Our network
model including the nearest neighbors and random SW links can be
seen in Fig.~\ref{fig_1dmodel}(b). Note that the occasional extra
checking of the simulated time of the random neighbor is
\textit{not} needed for the faithfulness of the simulation \cite{KORN03_SCI}.
It is merely introduced {\em to control} the width of the time horizon. The
occasional checking of the virtual time of the random neighbor
(with rate $p$) introduces an effective strength $J$$=$$J(p)$ for
these links. Note that this is a dynamic ``averaging" process
controlled by the parameter $p$ and can possibly be affected by
nonlinearities in the dynamics through renormalization effects.
The exact form of $J(p)$ is not known. The only plausible
assumption we make for $J$ is that it is a monotonically
increasing function of $p$ and is only zero when $p$=$0$.

In what follows, we focus on the characteristics of the  synchronization dynamics
on the network. As we have seen for the one-dimensional regular network, the
communication topology between the nodes (up to linear terms)
leads to simple relaxation, governed by the Laplacian on the
regular grid. Random communication links give rise to analogous
effective couplings between the nodes, corresponding to the
Laplacian on the random part of the network. Thus, the large-scale
properties of the virtual time horizon of our SW scheme are
governed by the effective Langevin equation
\begin{equation}
\partial_t \hat{\tau}_i = \nabla^2\hat{\tau}_i
- \sum_j J_{ij}(\hat{\tau}_i-\hat{\tau}_j) + ... + \eta_i(t) \;,
\label{meaf_KPZ}
\end{equation}
where the ... stands for infinitely many non-linear terms
(involving non-linear interactions through the random links as
well), and $J_{ij}$ is proportional to the symmetric adjacency matrix
of the random part of the network: $J_{ij}$$=$$J(p)$ if sites $i$ and
$j$ are connected by a random link and $J_{ij}$$=$$0$ otherwise.
For our specific SW construction each node has exactly one random
neighbor, i.e. there are no fluctuations in the individual
connectivity (degree) of the nodes. Our simulations (to be
discussed below) indicate that when considering the large-scale
properties of the systems, the Laplacian on the random part of the
network generates an effective coupling $\gamma$ to the mean value of the fluctuations
in the synchronization landscape $\overline{\tau}$ \cite{KORN03_SCI}.
At the level of the structure factor, it corresponds to an effective mass $\gamma$
(in a field-theory sense)
\begin{equation}
S(k)\propto\frac{1}{\gamma + k^2}\;,
\label{S_k}
\end{equation}
where $\gamma$$=$$\gamma(p)$ is a monotonically increasing
function of $p$ with $\gamma(0)$=$0$.

We emphasize that the above is not a derivation of
Eq.~(\ref{S_k}), but rather a ``phenomenological" description of
our findings. It is also strongly supported by exact
asymptotic results for the (linear) EW model on SW networks, where
the effect of the Laplacian on the random part of the network is
to generate an effective mass \cite{KOZMA03,KOZMA05b}. The averaging over the
quenched network ensemble, however, can introduce nontrivial
scaling and corrections in the effective coupling
\cite{KOZMA03,KOZMA05b,KOZMA05}. In our case, this is further
complicated by the nonlinear nature of the interaction. The
results of ``simulating the simulation'', however, suggest that
the dynamic control of the link strength and nonlinearities only
give rise to a renormalized coupling and a corresponding
renormalized mass. Thus, the dynamics of the BCS scheme with
random couplings is effectively governed by the EW relaxation in a
small-world \cite{KOZMA03,KOZMA05b,KOZMA05}.

From Eq.~(\ref{S_k}) it directly follows that the lateral
correlation length, in the infinite system-size limit, scales as
\begin{equation}
\xi\sim \gamma^{-1/2} \;,
\label{xi}
\end{equation}
i.e., becomes finite for all $p\not$=$0$. The presence of the
effective mass term in the structure factor in Eq.~(\ref{S_k})
implies that $\lim_{k\rightarrow 0}S(k)$$<$$\infty$, that is,
there are no large amplitude long-wavelength modes in the surface.
Consequently, the width $\langle w^2\rangle$$=$$(1/N)\sum_{k \not=
0}S(k)\sim\xi$ is also finite. Our simulated time landscapes
indeed show that they become macroscopically smooth when SW links
are employed [Figs.~\ref{figsn}(b) and \ref{figsn}(c)], compared
to the the same dynamics with only short-range links
[Fig.~\ref{figsn}(a)].
\begin{figure}[htb]
\vspace{6cm}
\includegraphics{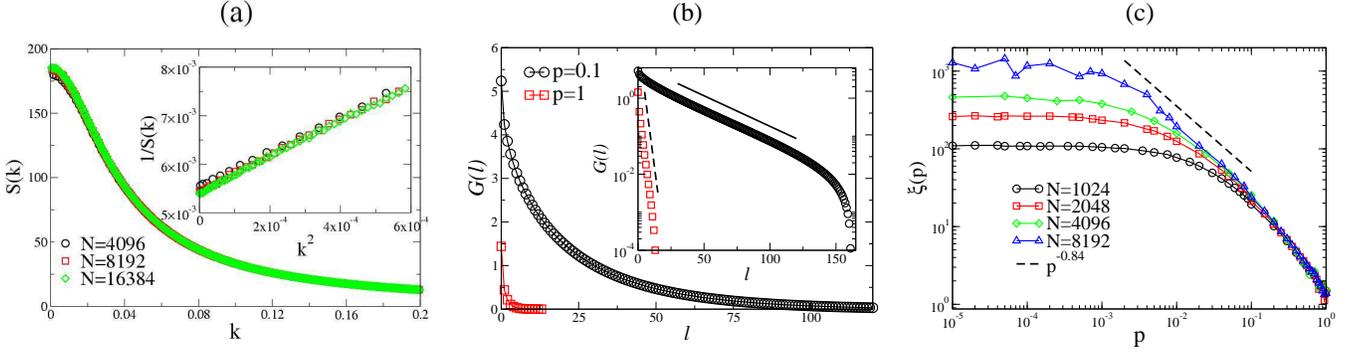}
\includegraphics{1D-QRM-corr.eps}
\includegraphics{1D-QRM-xi-p.eps}
\caption{(a) Structure factor for the 1D SW synchronization scheme with
$p$=$0.1$. The inset shows $1/S(k)$ vs. $k^2$ for small values of $k$, confirming the
coarse-grained prediction Eq.~(\ref{S_k}).
(b) The spatial two-point correlation function as a function of (the Euclidean) distance $l$
between the nodes for two different values of $p$, indicating an exponential decay with
an average correlation length $\xi$$\approx$$27$ and $\xi$$\approx$$1.6$ for $p$=$0.1$ and $p$=$1$,
respectively. The inset shows the same data on log-normal scales.
(c) Correlation length vs. $p$ for different system-sizes. The dashed line
corresponds to the estimate of the exponent $s$, $\xi(p)\sim p^{-s}$ with $s$$\approx$$0.84$,
in the small-$p$ regime for an asymptotically infinite system.}
\label{fig_sw1dsf}
\end{figure}

\begin{figure}[htb]
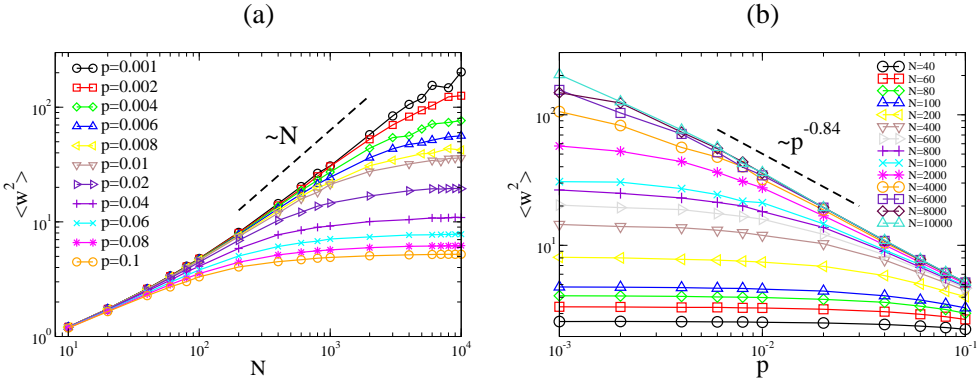

\vspace{6cm}
\includegraphics{1D-QRM-w2L-all.eps}
\includegraphics{1D-QRM-w2p-all.eps}
\caption{(a) The average steady-state width in the 1D SW synchronization landscape
as a function of the system size for different values of $p$ in the range of $[10^{-3},10^{-1}]$.
The dashed line indicates the EW/KPZ scaling, corresponding to the small system-size behavior.
(b) The average steady-state width in the 1D SW synchronization landscape
as a function of $p$ for different values of $N$ in the range of $[40,10^{4}]$.
The dashed line indicates the best-fit power law in the asymptotic large-$N$ small-$p$ regime
to extract the correlation length exponent $s$, according to
Eqs.~(\ref{new_scaling})-(\ref{xi_scale}).}
\label{fig_1d-w2-scaling}
\end{figure}

\begin{figure}[htb]
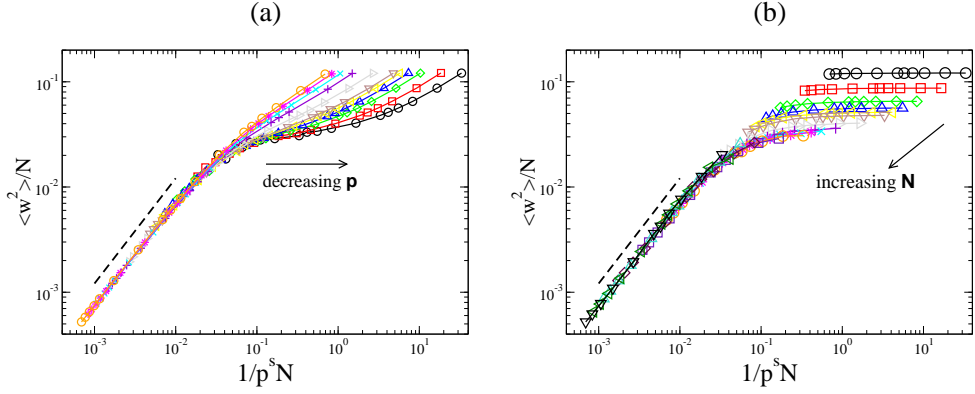

\vspace{6cm}
\includegraphics{1D-QRM-w2L-all-scaled.eps}
\includegraphics{1D-QRM-w2p-all-scaled.eps}
\caption{The scaled versions of Figs.~\ref{fig_1d-w2-scaling}(a) and (b), as proposed by
Eqs.~(\ref{new_scaling})-(\ref{xi_scale}) by plotting
$\left\langle w^2\right\rangle/N$ vs. $1/p^sN$ with $s$=$0.84$.
The data points are identical in (a) and (b),
but data points connected by a line
have the same value of $p$ in (a) and
have the same value of $N$ in (b), as obtained by rescaling
Figs.~\ref{fig_1d-w2-scaling}(b) and (b), respectively.
The dashed line corresponds to the asymptotic small-$x$ behavior of the scaling function $g(x)$
[Eq.~(\ref{f_newscaling})].}
\label{fig_1d-w2-scaling2}
\end{figure}

In the simulations, we typically performed averages over 10-100 network realizations,
and compared the results to those of individual runs. Our results indicate
that the observables we studied (the width and its distribution,
the structure factor, and the utilization) display strong {\em self-averaging} properties, i.e.,
for large enough systems, they become independent of the particular realization of the
underlying SW network. Simulation results for the structure factor, $S(k)$, for the SW
synchronization scheme are shown in Fig.~\ref{fig_sw1dsf}(a). If an
infinitesimally small $p$ is chosen, $S(k)$ approaches a finite
constant in the limit of $k$$\rightarrow$$0$, and in turn, the
virtual time horizon becomes macroscopically smooth with a
\textit{finite} width.

A possible (phenomenological) way to obtain the correlation
length is to fit our structure factor data to Eq.~(\ref{S_k}),
more specifically, by plotting $1/S(k)$ versus $k^2$, which exhibits a
linear relationship. By a linear fit, $\gamma$ is then the ratio
of the intercept and the slope (inset in
Fig.~\ref{fig_sw1dsf}(a)). Alternatively, one can confirm that the
massive propagator in Eq.~(\ref{S_k}) indeed leads to an exponential
decay in the two-point correlation function $G(l)$ from which the
correlation length can also be extracted
[Fig.~\ref{fig_sw1dsf}(b)]. In our case with a system-size
$N$=$16384$ it is $\xi$$\approx$$27$ for $p$=$0.1$ and
$\xi$$\approx$$1.6$ for $p$=$1$. Fig.~\ref{fig_sw1dsf}(c) shows
the correlation length extracted from the structure factor $S(k)$
as a function of $p$ for different system-sizes.

\begin{figure}[htb]
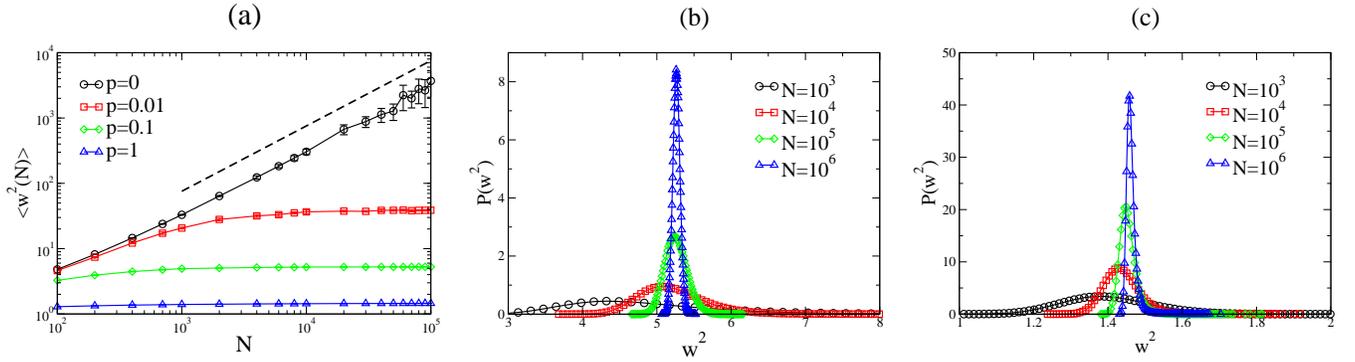

\vspace{5cm}
\includegraphics{1D-QRM-Pall-w2L.eps}
\includegraphics{1D-QRM-P0.10-w2-dist.eps}
\includegraphics{1D-QRM-P1.00-w2-dist.eps}
\caption{(a) The average steady-state width as a function of system-size for different values of $p$
in the 1D SW synchronization scheme. The $p$$=$$0$ case corresponds to the purely 1D BCS scheme
(exhibiting EW/KPZ scaling, indicated by a dashed line) and is also shown for
comparison. Steady-state width distributions in the 1D SW synchronization
scheme for (b) $p$=$0.1$ and for (c) $p$=$1$.
The distributions were constructed using ten different network realizations,
except for $N$=$10^6$, where only one realization was obtained due to
computational limitations. All width distributions, however,
indicated self-averaging.}
\label{fig_1d-kpz-qrm-w2dist}
\end{figure}

An alternative way to determine the correlation length is to
attempt a finite-size scaling analysis of the width $\langle
w^2\rangle$. There are two length scales in the system: the linear
system size $N$ and the correlation $\xi$ of an infinite system.
For $p$$=$$0$, $\langle w^2\rangle$$\sim$$N$, while for $p$$>$$0$
and $N$$\to$$\infty$, $\langle w^2\rangle$$\sim$$\xi$. For
non-zero $p$ and finite $N$ the scaling of the width can be
expected \cite{KOZMA05,KORN_PLA_swrn} to follow
\begin{equation}
\left\langle w^2\right\rangle = N g(\xi/N)\;,
\label{new_scaling}
\end{equation}
where $g(x)$ is a scaling function such that
\begin{equation}
g(x) \sim \left\{
\begin{array}{ll}
x    & \mbox{if $x$$\ll$$1$} \\
\mbox{const.} & \mbox{if $x$$\gg$$1$}
\end{array}
\right. \;.
\label{f_newscaling}
\end{equation}
For non-zero $p$ and for sufficiently small systems ($N$$\ll$$\xi(p)$)
one can confirm that the behavior of the width follows that of the
system without random links $\langle w^2\rangle\sim N$
[Fig.~\ref{fig_1d-w2-scaling}(a)]. For large-enough systems, on
the other hand, we can extract the $p$-dependence of the
infinite-system correlation length as $\langle w^2\rangle\sim
\xi(p)$ [Fig.~\ref{fig_1d-w2-scaling}(b)], yielding
\begin{equation}
\xi(p) \sim p^{-s} \;,
\label{xi_scale}
\end{equation}
where $s\approx0.84$.

\begin{figure}[htb]
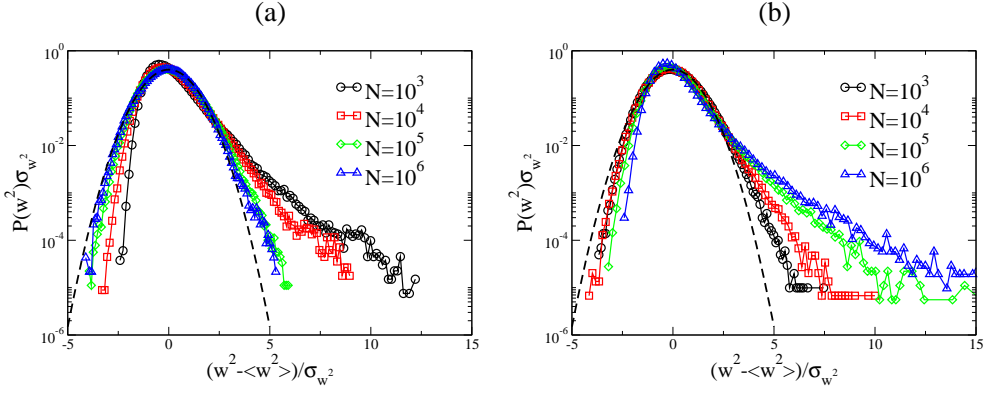

\vspace{5.5cm}
\includegraphics{1D-QRM-P0.10-w2-dist-scaled.eps}
\includegraphics{1D-QRM-P1.00-w2-dist-scaled.eps}
\caption{Steady-state width distributions for the 1D SW synchronization scheme scaled to zero mean
and unit variance for (a) $p$=$0.1$ and (b) $p$=$1$.  The dashed curves are similarly scaled
Gaussians for comparison.}
\label{fig_1d-qrm-w2dist-scaled}
\end{figure}

We then studied the data collapse as proposed by
Eq.~(\ref{new_scaling}) by plotting $\langle w^2\rangle/N$ vs.
$1/p^{s}N$. In fact, we performed this rescaling
originating from both raw data sets
Figs.~\ref{fig_1d-w2-scaling}(a) and (b). The resulting scaled
data points in Fig.~\ref{fig_1d-w2-scaling2}(a) and (b), of course, are
identical in the two figures, but the lines connect data points
with the same value of $p$ in Fig.~\ref{fig_1d-w2-scaling2}(a) and
with the same value of $N$ in Fig.~\ref{fig_1d-w2-scaling2}(b).
These scaled plots in Fig.~\ref{fig_1d-w2-scaling2} indicate
that there are very strong corrections to scaling: data for larger
$p$ or smaller $N$ values peel off from the proposed scaling form
Eq.~(\ref{f_newscaling}) relatively quickly. These strong corrections are
possibly the result of the nonlinear nature of the interaction
between the nodes on the quenched network. We note that the
purely linear EW model on identical networks exhibits the scaling
proposed in Eqs.~(\ref{new_scaling}) and (\ref{f_newscaling})
{\em without} noticeable corrections to scaling \cite{KOZMA05,KORN_PLA_swrn}.

The non-zero $\gamma$, leading to a finite correlation length,
$\xi$, ensures a finite width in the infinite system-size limit.
Our simulations show that the width saturates to a finite value
for $p$$>$$0$ [Fig.~\ref{fig_1d-kpz-qrm-w2dist}(a)]. The
distribution of the steady-state width $P(w^2)$ changes from that
of the EW/KPZ class to a delta function for non-zero values of $p$
as the system size goes to infinity.
Figure~\ref{fig_1d-kpz-qrm-w2dist}(b) and (c) shows the width
distributions for $p$=$0.1$ and $p$=$1$, respectively. The scaled
width distributions (to zero mean and unit variance), however,
exhibit the convergence to a delta function through nontrivial
shapes for different values of $p$. For $p$=$0.1$
[Fig.~\ref{fig_1d-qrm-w2dist-scaled}(a)] the distributions appear
to slowly converge to a Gaussian as the system-size increases. For
$p$=$1$  [Fig.~\ref{fig_1d-qrm-w2dist-scaled}(b)], the trend is
\emph{opposite}, up to the system sizes we could simulate; as the
system-size increases, the distributions exhibit progressively
non-Gaussian features (closer to an exponential) around the
center, for up to $N$=$10^6$. Note that not only the average width
$\langle w^2\rangle$, but also the full distribution $P(w^2)$ was
found to be \emph{self-averaging}, i.e., independent of the
particular realization of the underlying SW network, so the above
effect is not due to insufficient averaging over network
realizations.
Since we were intrigued by the above change in the trend of the
convergence (i.e., away from vs. toward the Gaussian), we carried
out some exploratory simulations for a range of $p$ values and
evaluated the kurtosis and the skewness of the width
distributions. The results shows that this change occurs at around
$p\approx 0.9$. We could not conclude, however, whether this
change corresponds to an actual transition and the emergence of a
strong disordered-coupling regime, where the distribution of the
width is a delta function, centered about a finite value, but, the
limiting shape of the delta function is \emph{non-Gaussian}, or to
the development of strong non-monotonic finite-size effects.
\begin{figure}[htb]
\vspace{6.2cm}
\includegraphics{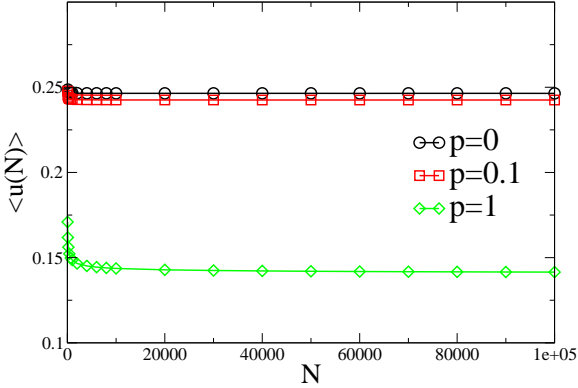}
\vspace{-0.5cm}
\caption{Steady-state utilization of SW synchronization network in 1D as a function of
system-size for three different values of $p$$=$$0$ (BCS), $p$$=$$0.1$ and $p$$=$$1$.}
\label{fig_1d-qrm-util}
\end{figure}

The effect of the random communication links on the utilization
can be understood as follows. According to the algorithmic rules,
the virtual times of the full network neighborhood (including the
random neighbor) are checked with probability $p$, while with
probability $(1$$-$$p)$ only short-ranged synchronization is employed.
Thus, the average progress rate of the simulated times becomes
\begin{eqnarray}
\langle u \rangle & = & (1-p) \langle\Theta(-\phi_{i-1})
\Theta(\phi_{i}) \rangle + p \langle\Theta(-\phi_{i-1})
\Theta(\phi_{i}) \Theta(\tau_{r(i)}-\tau_{i}) \rangle \nonumber \\
& = &
\langle\Theta(-\phi_{i-1}) \Theta(\phi_{i}) \rangle -
p \left[ \langle\Theta(-\phi_{i-1}) \Theta(\phi_{i}) \rangle -
\langle\Theta(-\phi_{i-1}) \Theta(\phi_{i}) \Theta(\tau_{r(i)}-\tau_{i})
\rangle \right] \;.
\label{u_SW}
\end{eqnarray}
Note that performing disorder averaging (over random network
realizations) makes the right hand side independent of $i$. In the
presence of the SW links the regular density of local minima
$\langle\Theta(-\phi_{i-1}) \Theta(\phi_{i}) \rangle$ {\em remains
nonzero} (in fact, increases, compared to the short-range
synchronized BCS scheme) \cite{GUCLU_CNLS,TORO03,KORN03_SCI}.
Thus, for an infinitesimally small $p$, the utilization, at most,
can be reduced by an infinitesimal amount, and the SW-synchronized
simulation scheme maintains a nonzero average progress rate. This
can be favorable in PDES where scalable global performance
requires both finite width and nonzero utilization. With the SW
synchronization scheme, both of these objectives can be achieved.
For example, for $p$$=$$0.1$ $\langle u \rangle_{\infty} \simeq
0.242$, while for $p$$=$$1$ $\langle u \rangle_{\infty} \simeq
0.141$. The steady-state utilization as a function of system size
for various values of $p$ can be seen in
Fig.~\ref{fig_1d-qrm-util}.

\end{section}


\begin{section}{Basic Conservative Synchronization Scheme in 2D}

A natural generalization to pursue is the same synchronization dynamics
and the associated landscapes on networks embedded in higher
dimensions. One might ask whether PDES of two-dimensional
phenomena exhibit kinetic roughening of the virtual time horizon.
Preliminary results indicated that this is the case
\cite{KORN00_UGA, KIRKPATRICK03}. In this section we give detailed
results when the BCS scheme is extended onto a two-dimensional
lattice ($N_{PE}$$=$$N^2$, $N$ being the linear size)
in which each node has four nearest neighbors. We consider
a system with periodic boundary conditions in both axis as can be
seen in Fig.~\ref{fig_2d-model}(a). The same microscopic rules,
i.e. each node increments its local simulated time by an
exponentially distributed random amount when it is a local minima
among its nearest neighbors, are applied to this lattice. As in
the one-dimensional case, during the evolution of the local
simulated times correlations between the nodes develop in the system.
One observes rough time surfaces in the steady-state after
simulating this system. Figure~\ref{fig_2d-surface}(a) shows the
contour plot of the simulated time surface for the BCS scheme in 2D.
\begin{figure}[htb]
\vspace{5cm} \includegraphics{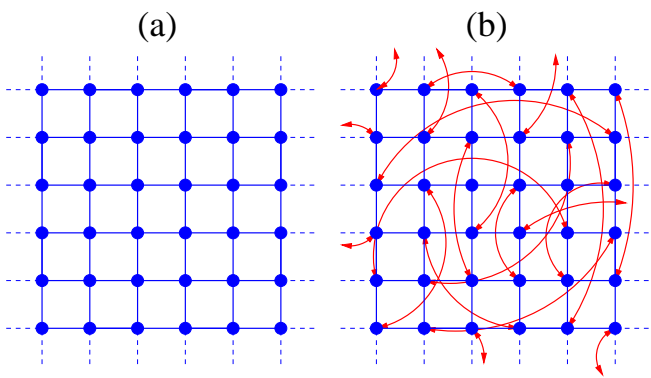} \caption{Communication
topologies in 2D. (a) 2D regular network, where each node is connected to its four nearest neighbors.
(b) SW synchronization network. One random link per node is added on top
of the 2D regular network. Red arrows show the bidirectional random links between the nodes.}
\label{fig_2d-model}
\end{figure}
\begin{figure}[htb]
\vspace{6cm}
\includegraphics{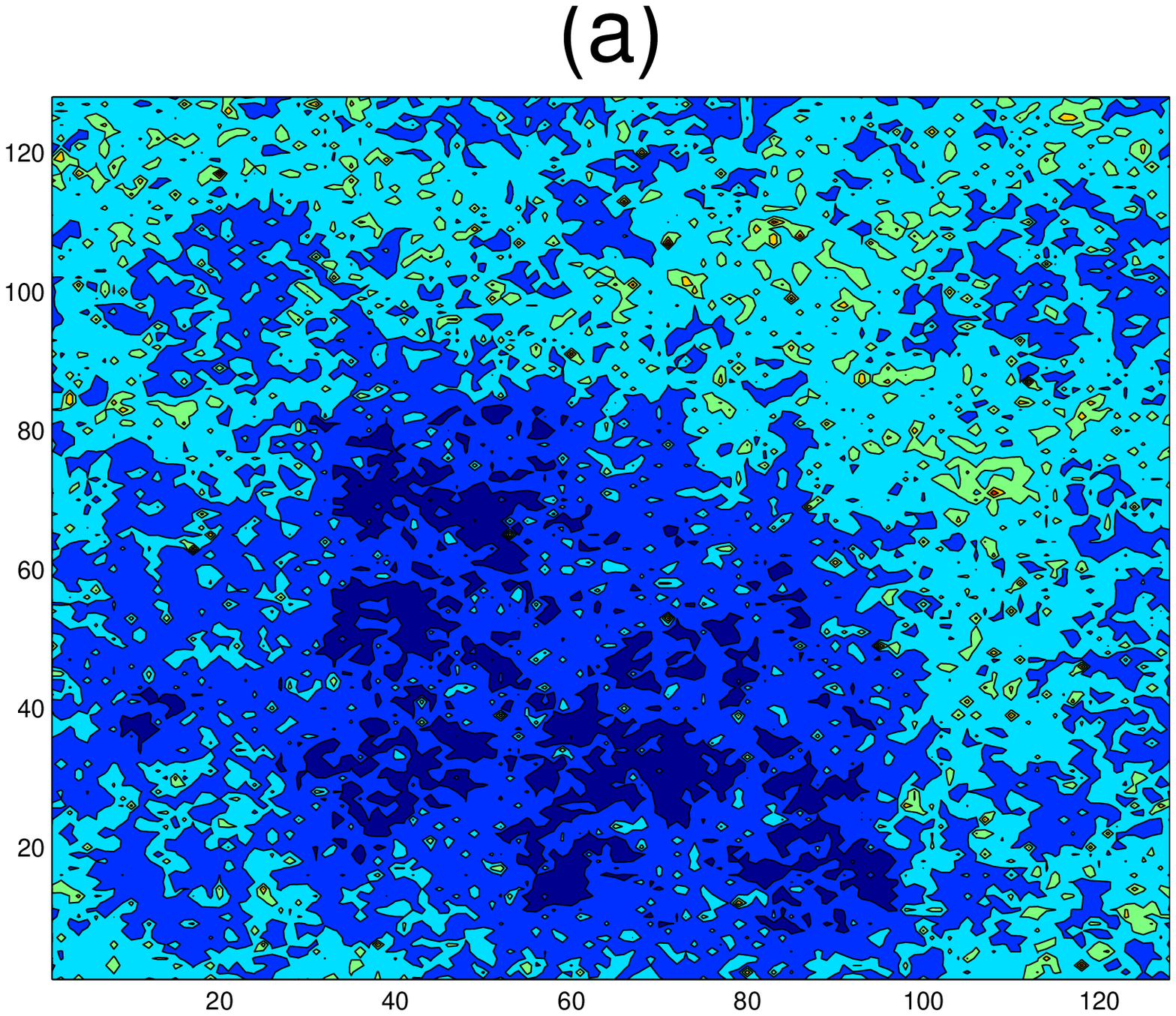}
\includegraphics{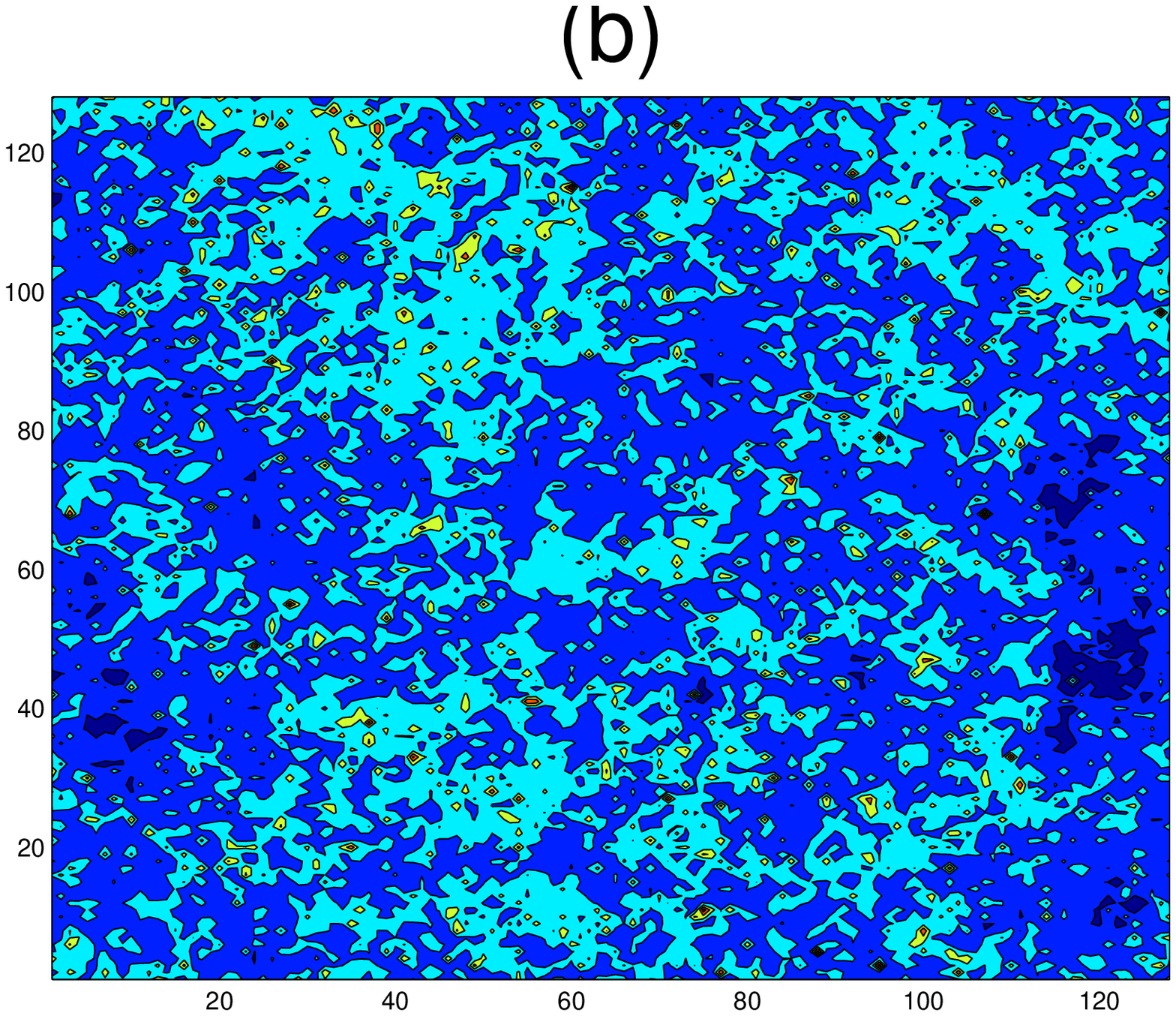}
\includegraphics{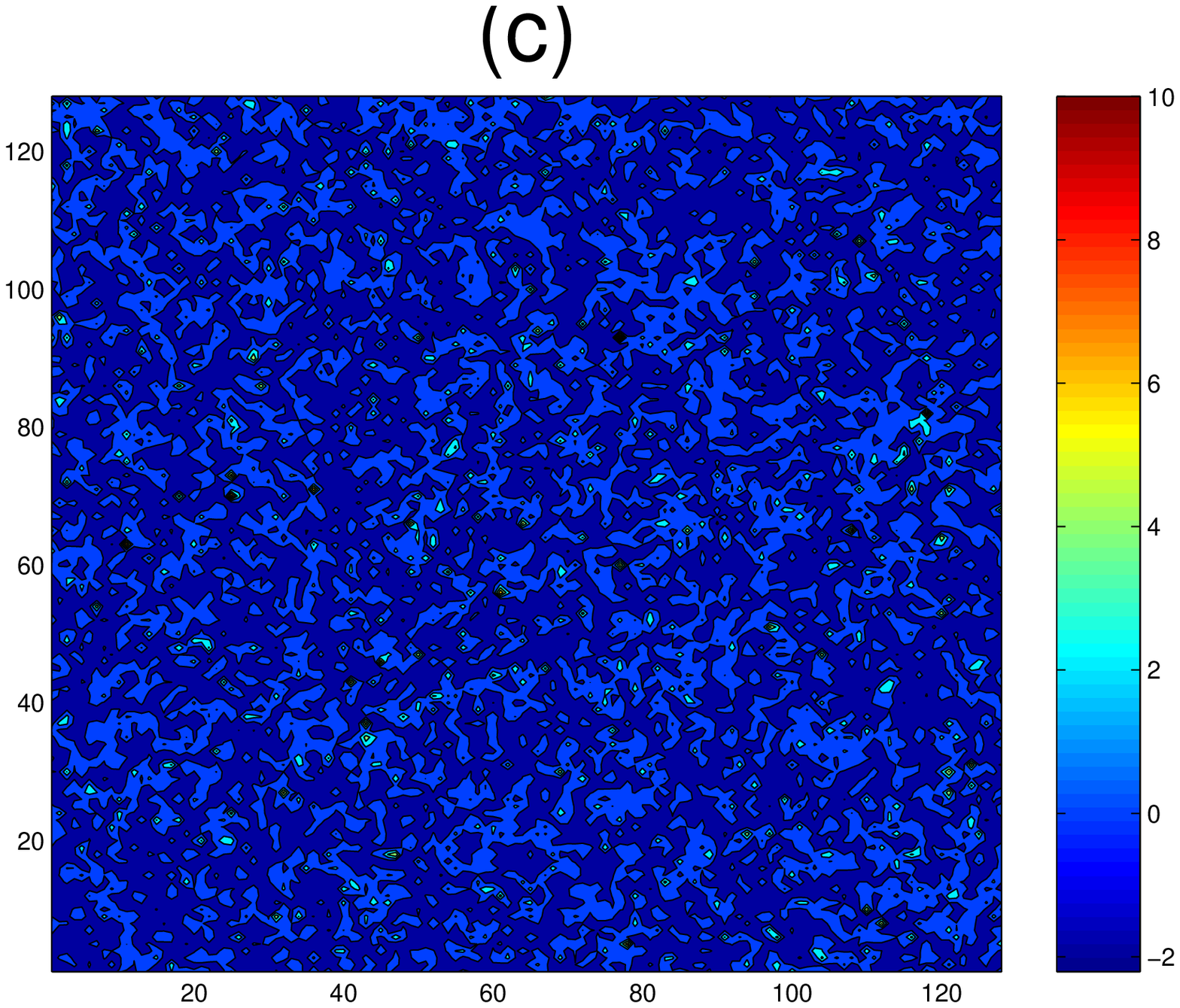}
\caption{Synchronization landscapes as contour plots for the 2D BCS on regular and SW
networks of 128x128 nodes. (a) for the BCS scheme on a regular lattice with only
nearest-neighbor connections (equivalent to $p$=$0$); (b) for $p$=$0.1$; (c) for $p$=$1$.}
\label{fig_2d-surface}
\end{figure}
In 2D as well, we observe kinetic roughening of the BCS scheme.
The simulated time surface for a finite system initially roughens with time
in a power-law fashion. It then saturates after some system-size
dependent crossover time to its system-size dependent steady-state
value, as shown in [Fig.~\ref{fig_2d-w2}(a)]. Our estimate for the
growth exponent in the early-time regime is $\beta$$=$$0.125$,
significantly smaller than in one dimension.

The roughness exponent $\alpha$ for KPZ-like systems have been
measured and estimated in a number of experiments and simulations
\cite{BARA95}. Since exact exponents for the higher-dimensional
KPZ universality class are not available, for reference, we
compare our results to a recent high-precision simulation study by
Marinari et al. \cite{MARINARI00} on the restricted solid-on-solid
(RSOS) model \cite{KIM89}, a model believed to belong to the KPZ
class. They found in Ref. \cite{MARINARI00} that $\alpha$$\simeq$$0.39$ for the 2D
RSOS roughness exponent. While our simulations of the virtual time
horizon show kinetic roughening in Fig.~\ref{fig_2d-w2}(a), the
scaled plot, suggested by Eq.~(\ref{vicsek}), indicates very
strong corrections to scaling  for the BCS in 2D [Fig.~\ref{fig_2d-w2}(a) inset]. Figure
~\ref{fig_2d-w2}(b) and (c) also indicates that the (KPZ) scaling
regime is approached very slowly, which is not completely
unexpected: for the 1D BCS scheme as well, convergence to the
steady-state roughness exponent Fig.~\ref{fig_kpz1dw2Lt}(a) and to
the KPZ width distribution Fig.~\ref{fig_kpz1dw2Lt}(b) only
appears for linear system sizes $N$$>$${\cal O}(10^3)$. Here, for the
2D case, the asymptotic roughness scaling [Fig.~\ref{fig_2d-w2}(b)]
and width distribution [Fig.~\ref{fig_2d-w2}(c)] has not been
reached for the system sizes we could simulate (up to linear
system size $N$$=$$2048$). Nevertheless the trend in the finite-size
behavior, and the identical microscopic rules (simply extended to
2D) suggest that 2D BCS landscape belongs to the 2D KPZ universality
class.
\begin{figure}[htb]
\vspace{5cm}
\includegraphics{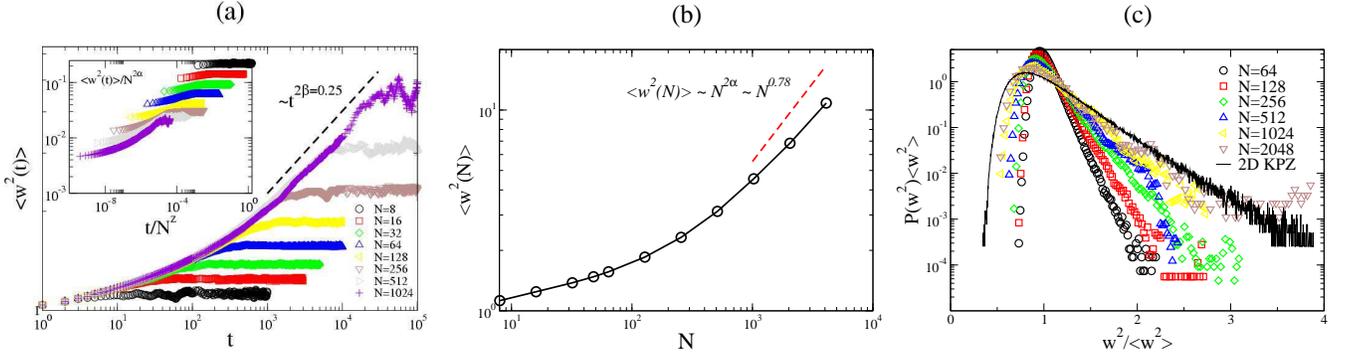}
\includegraphics{2D-KPZ-w2L.eps}
\includegraphics{2D-KPZ-w2-dist-scaled.eps}
\caption{(a) Time evolution of the width in 2D BCS scheme. The
dashed line indicates the power-law behavior of the width before saturation with a
growth exponent $\beta$$\approx$$0.125$. The inset shows the scaled plot
$\left\langle w^2\right\rangle/N^{2\alpha}$ vs. $t/N^z$.
(b) Steady-state width of the 2D BCS scheme as a function of the linear system-size.
The dashed line corresponds to the asymptotic 2D KPZ scaling with roughness exponent
$2\alpha$$=$$0.78$ as obtained by high-precision simulations of the RSOS model
\cite{MARINARI00}. Note the log-log scales.
(c) The scaled width distributions for the 2D BCS scheme.
The solid curve is the asymptotic 2D KPZ scaled width distribution, again from
high-precision RSOS simulations \cite{MARINARI02}. Note the log-normal scales.}
\label{fig_2d-w2}
\end{figure}

For further evidence, we also constructed the structure factor for the 2D BCS
steady-state landscape. As shown in Fig.~\ref{fig_2d-kpz-sf}(a),
$S(k_x,k_y)$ exhibits a strong singularity about ${\bf k}={\bf
0}$. For further analysis, we exploited the symmetry of
$S(k_x,k_y)$ that it can only depend on $|{\bf k}|=\sqrt{k_{x}^2 +
k_{y}^2}$. Hence, we averaged over all directions having the same
wave number $|{\bf k}|$ to obtain $S(|{\bf k}|)$.  For small wave
numbers we found that it diverges as
\begin{equation}
S(|{\bf k}|) \sim \frac{1}{|{\bf k}|^{2+2\alpha}} \;,
\label{S_k_2D_KPZ}
\end{equation}
with $\alpha=0.39$, as shown in Fig.~\ref{fig_2d-kpz-sf}(b). This
is consistent with the small-$k$ behavior of the structure factor
of the 2D KPZ universality class with roughness exponent
$\alpha$$\simeq$$0.39$ \cite{MARINARI00}. As noted above, the scaling of the width
and its distribution exhibited very slow convergence to those of our ``reference"-KPZ system, the
RSOS model \cite{MARINARI00,MARINARI02}. This is likely the effect of the non-universal and
surprisingly large contributions coming from the large-$k$ modes, leading to very strong
corrections to scaling for the system sizes we were able to study in 2D.
Looking directly at the small-$|{\bf k}|$ behavior of $S(|{\bf k}|)$ is, of course,
undisturbed by the larger-$|{\bf k}|$ modes,
hence the relatively good agreement with the 2D KPZ scaling.

\begin{figure}[htb]
\vspace{6cm} \includegraphics{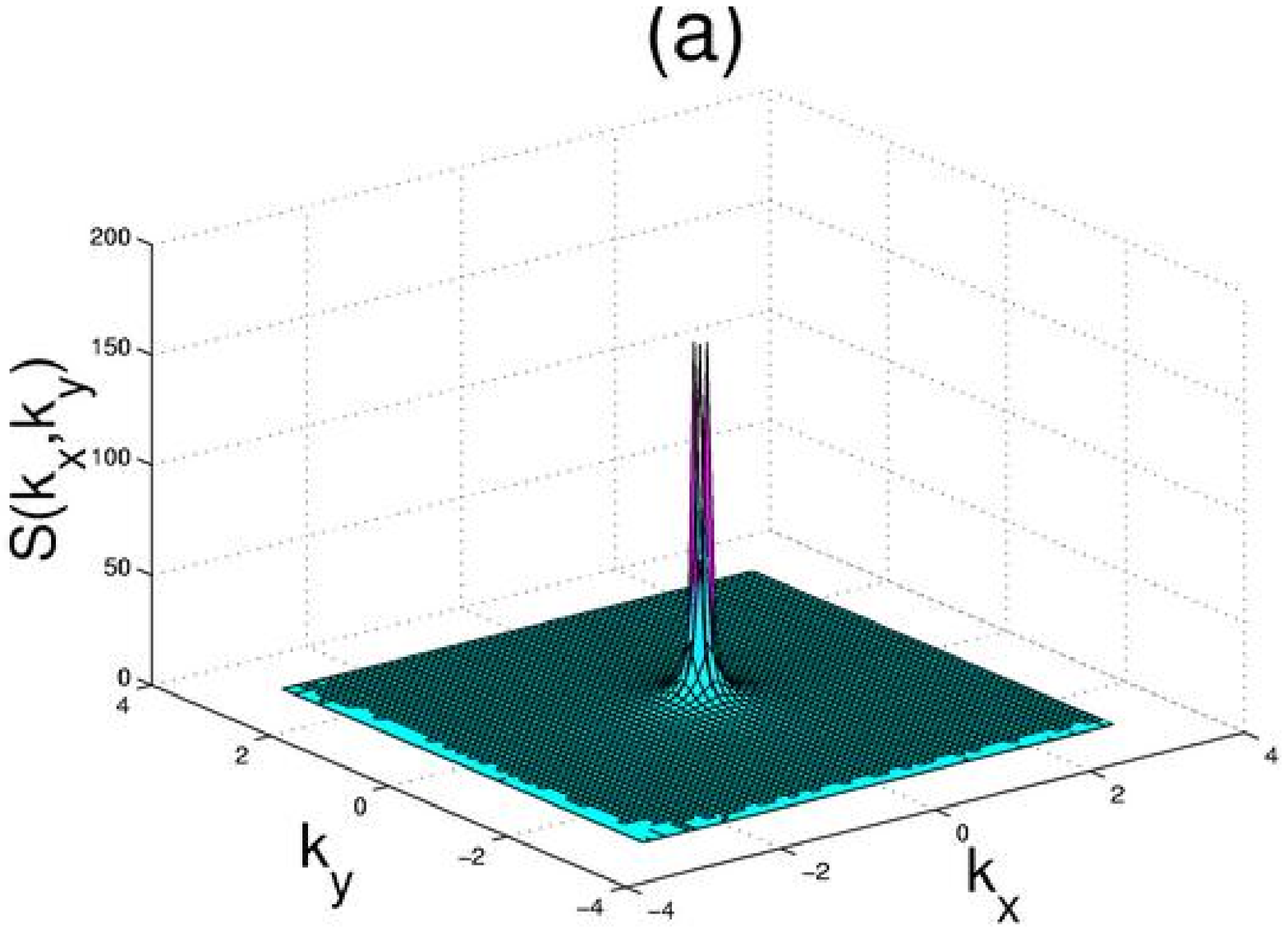}
\includegraphics{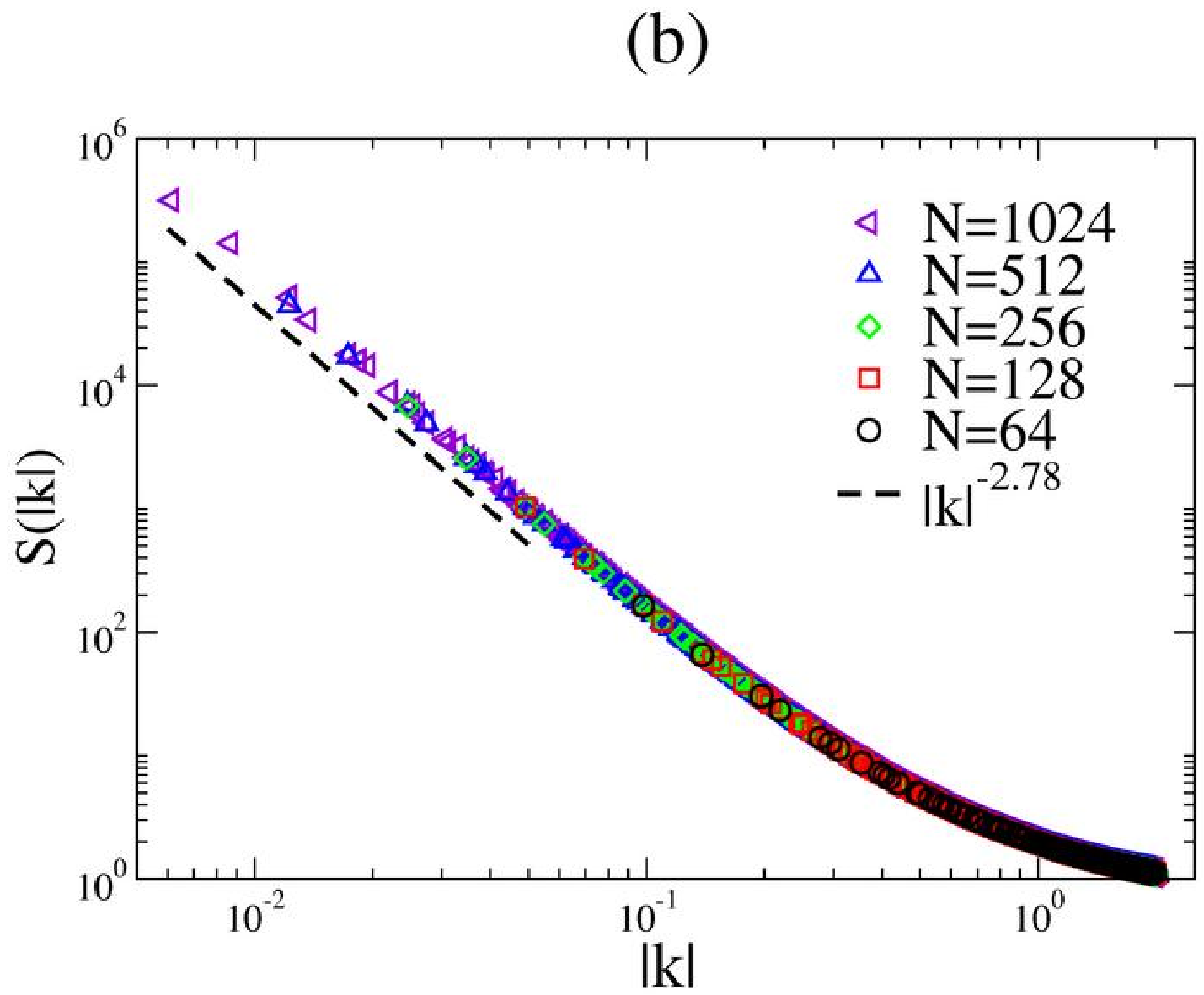}
\caption{Structure factor for the BCS scheme in 2D.
(a) as a function of the wave number components $k_x$ and $k_y$.
(b) as a function of the magnitude of the wave number for
different system-sizes. The dashed line shows the asymptotic 2D KPZ behavior
for small values of $|{\bf k}|$ [Eq.~(\ref{S_k_2D_KPZ})] with $\alpha$$=$$0.39$ \cite{MARINARI00}.
Note the log-log scales.}
\label{fig_2d-kpz-sf}
\end{figure}

\begin{figure}[htbp]
\vspace{6cm}
\includegraphics{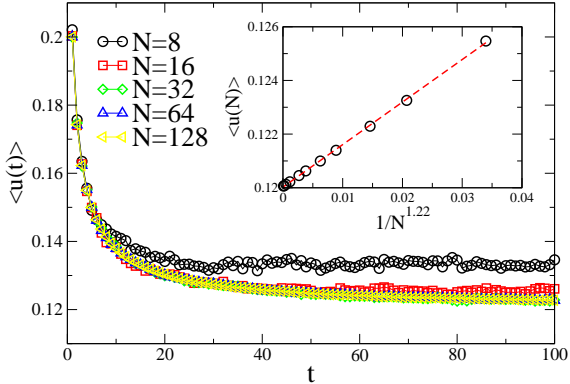}
\vspace{-0.5cm}
\caption{The time evolution of the steady-state utilization in 2D BCS scheme
for various system sizes.
The inset shows the steady-state utilization in the 2D BCS scheme as a function of
$1/N^{2(1-\alpha)}$ with the 2D KPZ
roughness exponent $\alpha$$=$$0.39$.
The dashed line is a linear fit to
Eq.~(\ref{utild}) with $\alpha$$=$$0.39$,
$\langle u(N)\rangle \approx 0.1201 + 0.1585/N^{1.22}$.}
\vspace{-0.5cm}
\label{fig_2d-kpzutil}
\end{figure}

The steady-state utilization (density of local minima) in the 2D
BCS synchronization landscape approaches a nonzero value in the
limit of an infinite number of nodes, $\langle u(\infty) \rangle$$\simeq$$0.1201$
as can be seen in Fig.~\ref{fig_2d-kpzutil}.
This is consistent with the general {\em approximate}
behavior $\langle u(\infty) \rangle\simeq const./d$ on hypercubic
lattices in $d$ dimension \cite{GREEN96,KORN00_UGA}, i.e.,
$\langle u(\infty) \rangle$ is approximately inversely proportional to the
coordination number. As shown
in Fig.~\ref{fig_2d-kpzutil} (inset), for a two-dimensional BCS scheme,
the steady-state utilization scales as [Eq.~(\ref{utild})]
$\langle u(N)\rangle \simeq \langle u(\infty)\rangle + \frac{const.}{N^{1.22}}$,
where we have used the 2D KPZ roughness exponent $\alpha$$=$$0.39$ \cite{MARINARI00}.

\end{section}


\begin{section}{Small-World Synchronization Scheme in 2D}

The de-synchronization (roughening of the virtual time horizon)
again motivates the introduction of the possibly long-range,
quenched random communication links on top of the 2D regular
network. Each node has exactly one (bi-directional) random
link as illustrated in Fig.~\ref{fig_2d-model}(b). The
actual ``microscopic" rules are analogous to the 1D SW case: with
probability $p$ each node will check the local simulated times of
all of its neighbors, including the random one, and can increment
its local simulated time by an exponentially distributed random
amount, only if it is a ``local" minimum (among the four
nearest neighbors and its random neighbor). With probability
$(1$$-$$p)$, only the four regular lattice neighbors are checked for
the local minimum condition.

The effect of the synchronization through the random links is,
again, to stop kinetic roughening and to suppress fluctuations in
the synchronization landscapes. Contour plots of the
synchronization landscapes are shown in
Figure~\ref{fig_2d-surface}(b) and (c) for $p$$=$$0.1$ and $p$$=$$1$,
respectively. Our results indicate that for any nonzero $p$ the
width of the surface approaches a finite value in the limit of
$N$$\to$$\infty$ [Fig.~\ref{fig_2d-qrm-w2}(a)]. At the same time,
the width distribution approaches a delta-function in the large
system-size limit as shown in Figs.~\ref{fig_2d-qrm-w2}(b) and \ref{fig_2d-qrm-w2}(c).
The scaled distributions (to zero mean and unit variance) again
show that at least for the finite systems we observed, the shape
of these distribution differs from a Gaussian. The deviation from
the Gaussian around the center of the distribution is stronger for
a larger value of $p$ where the influence of the quenched random
links are stronger. Note that for the 1D SW landscapes as well,
the width distribution only displayed a crossover to Gaussian
behavior for smaller values of $p$ and for very large linear
system sizes ($N$$>$${\cal O}(10^4)$. In the 2D SW case, these linear
system sizes are computationally not attainable, and the
convergence to a Gaussian width distribution, or the existence of an inherently
different strong-disorder regime (as a result of strong random links),
remains an open question.

The underlying reason for the finite width is again a finite
average correlation length between the nodes. The 2D structure
factor exhibits a massive behavior, i.e., $S(|{\bf k}|)$
approaches a finite value in the limit of ${\bf k}$$\to$${\bf 0}$
[Figs.~\ref{fig_2d-sf}(a) and \ref{fig_2d-sf}(b)]. For small wave numbers, the
approximate behavior of the structure factor is again
$S(|{\bf k}|)$$\propto$$1/(|{\bf k}|^2+\gamma)$ as can be seen in
the inset of Fig.~\ref{fig_2d-sf}(b), with strong finite-size corrections to $\gamma$.
The relevant feature of the synchronization dynamics on a SW network is the
generation of the effective mass $\gamma$. Nonlinearities can give
rise to a renormalized mass, but the relevant operator is the
Laplacian on the random network.

In the 2D SW synchronization scheme the steady-state utilization
is smaller than its purely 2D counterpart (BCS in 2D), as a result of the
possible additional checking with the random neighbors. For small
values of $p$, however, it is reduced only by a small amount, and
remains nonzero in the limit of an infinite number of nodes
[Fig.~\ref{fig_2d-qrm-util}]. For
example, for $p$$=$$0.1$ $\langle u
\rangle_{\infty}$$\simeq$$0.1198$, while for $p$=$1$ $\langle u
\rangle_{\infty}$$\simeq$$0.084$.
\begin{figure}[htb]
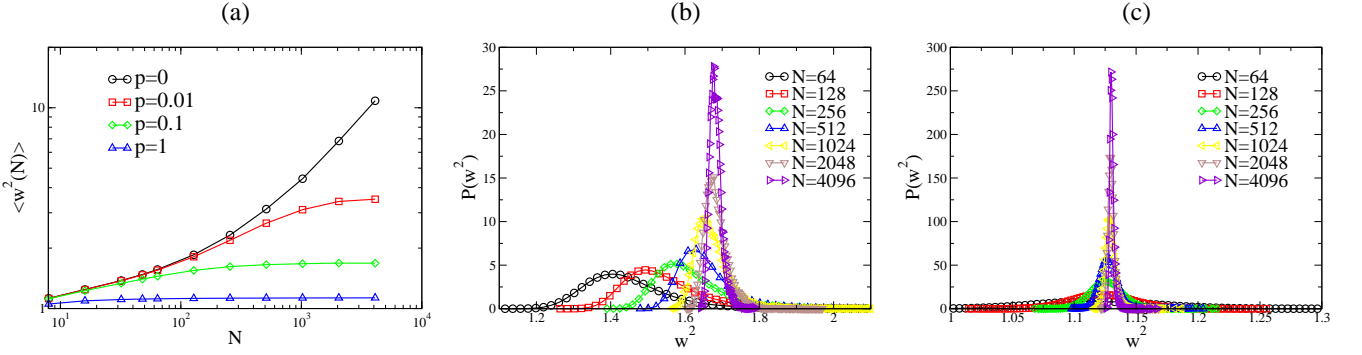

\vspace{5cm}
\includegraphics{2D-QRM-w2L.eps}
\includegraphics{2D-QRM-P0.10-w2-dist.eps}
\includegraphics{2D-QRM-P1.00-w2-dist.eps}
\caption{(a) The average steady-state width as a function of linear system size for different
$p$ values in the 2D SW synchronization scheme.
The data for $p$=$0$ corresponds to the 2D BCS
scheme on a regular network with only nearest-neighbor connections.
Steady-state width distributions in the 2D SW synchronization scheme
for (b) $p$$=$$0.1$ and for (c) $p$$=$$1$ for various system sizes.}
\label{fig_2d-qrm-w2}
\end{figure}
\begin{figure}[htb]
\vspace{6cm}
\includegraphics{2D-QRM-P0.10-w2-dist-scaled.eps}
\includegraphics{2D-QRM-P1.00-w2-dist-scaled.eps}
\caption{Steady-state width distributions in the 2D SW synchronization scheme
scaled to zero mean and unit variance
for (a) $p$$=$$0.1$ and for (b) $p$$=$$1$.
The dashed curves are similarly scaled Gaussians for comparison.}
\label{fig_2d-qrm-w2dist-scaled}
\end{figure}

\begin{figure}[htb]
\vspace{5cm}
\includegraphics{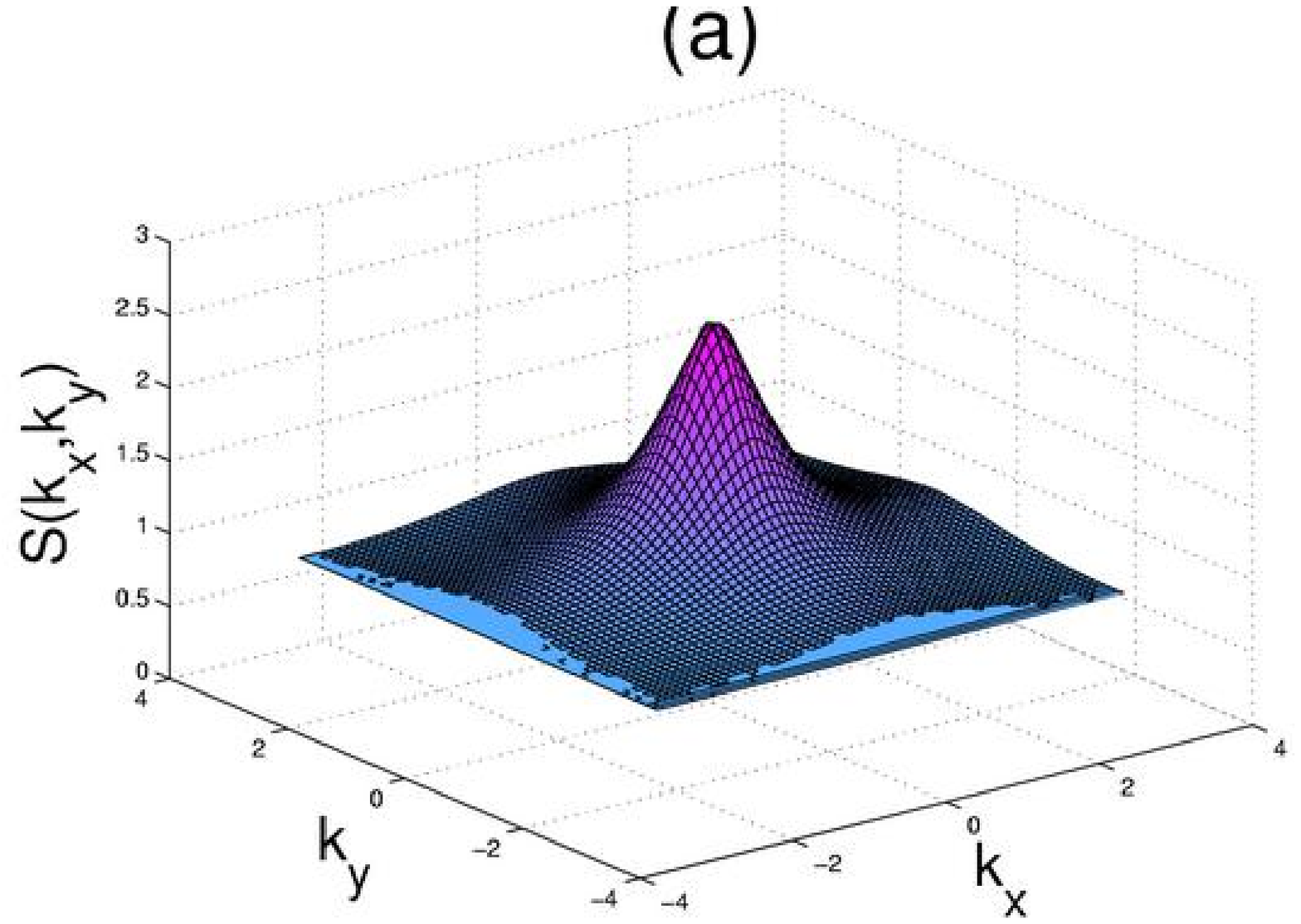}
\includegraphics{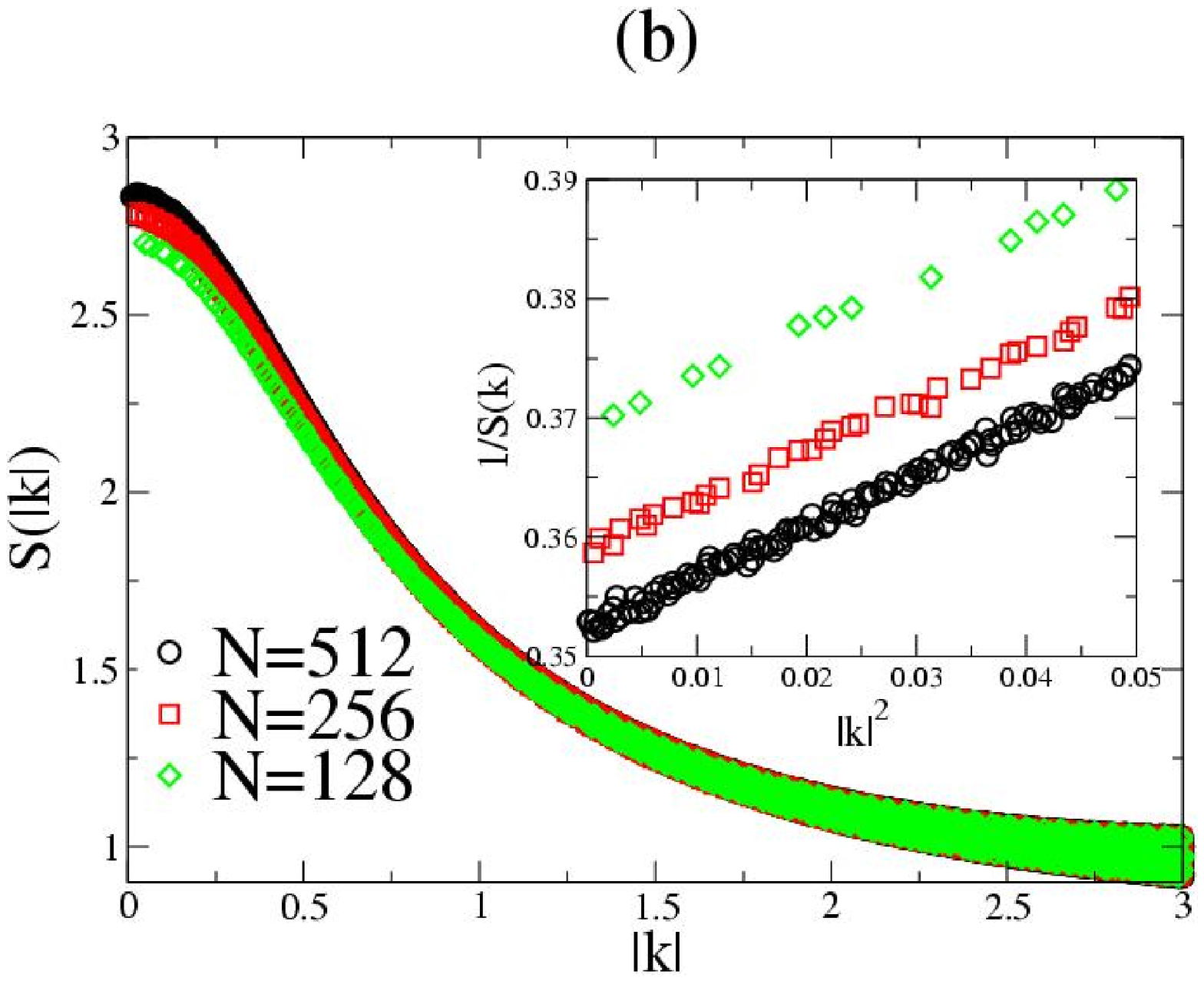}
\caption{(a) Steady-state structure factor for the 2D SW scheme
as a function of the components $k_x$ and $k_y$ for $p$$=$$1$.
(b) Structure factor as a function of $|\vec{k}|$ for the 2D SW synchronization scheme
for $p$=$1$. The inset shows $1/S(k)$ vs. $k^2$ for small values of $|k|$.}
\label{fig_2d-sf}
\end{figure}


\begin{figure}[htb]
\vspace{7cm}
\includegraphics{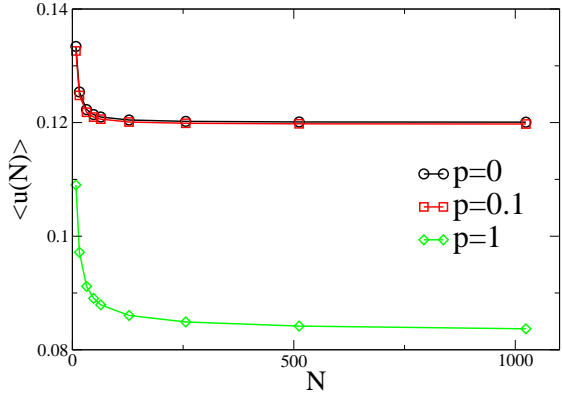}
\vspace{-0.5cm}
\caption{Steady-state utilization of SW synchronization network in 2D
as a function of system-size for three different
values of $p$$=$$0$ (BCS), $p$$=$$0.1$ and $p$$=$$1$.}
\vspace{-0.5cm}
\label{fig_2d-qrm-util}
\end{figure}

\end{section}


\begin{section}{Summary and Conclusions}

We studied the large-scale properties of the synchronization
landscapes of PDES, applicable to certain distributed and
networked computer systems for a large number of scientific
problems. We investigated the effects of SW-like additional
communication links between the nodes added on the top of 1D and
2D regular networks. With purely regular short-range connections
the synchronization landscape is rough and belongs to the KPZ
universality class. This property can hinder efficient data
management. To suppress diverging fluctuations (as a function of
the number of the nodes) we constructed a SW synchronization
scheme where the nodes also synchronize with a randomly chosen one
with a possibly infinitesimal frequency. For an infinitesimally
small rate of communications over the random links, this scheme
results in a progress rate reduced only by an infinitesimal
amount, while the width of the time horizon becomes \emph{finite}
in the limit of infinitely many PEs. Thus, the scheme is fully
scalable as the PEs make nonzero and close-to-uniform progress
without global intervention. In obtaining our results, we used
coarse-grained arguments to identify the universality class of the
PDES time horizon, and confirmed those predictions by actually
simulating the simulations, based on the exact algorithmic rules.
Our study of the width distributions of SW-synchronized PDES
landscapes also revealed that while they converge to
delta-functions, centered about a \emph{finite value}, their shape
progressively becomes \emph{non-Gaussian} for larger values of
$p$. We could not conclude if these observations indicate the
emergence of a strong-disorder regime, or strong non-monotonic
finite size effects for our attainable system sizes. Future work
may address these questions.

In this work we considered the simplest (and in some regards, the
worst case) scenario, where each nodes carries one site of the
underlying physical system, hence synchronization with nearest
neighbor PEs is required at every step. In actual parallel implementations the
efficiency can be greatly increased by hosting many sites by each
PE \cite{KORN99_JCP,AMAR_PRB_2005a,AMAR_PRB_2005b}.
That way, communication between PEs is only required when
local variables are to be updated on the boundary region of the
sites hosted by the PEs (within the finite range of the
interactions). While the above procedure clearly increases the
utilization and reduces the actual communication overhead, it
gives rise to an even faster growing early time regime in the
simulated time horizon \cite{KOLA_PRE_2004}. Since the PEs rarely need to
synchronize, up to some crossover time, the evolution of the time
horizon is governed by random deposition \cite{BARA95}, a faster
roughening growth, before eventually crossing over to the KPZ
growth and a subsequent saturation.

Our findings are also closely related to critical phenomena and
collective phenomena on networks
\cite{ALBERT02,DOROGOVTSEV02,STROGATZ01,GOLTSEV03}. In particular,
in recent years, a number of prototypical models have been
investigated on SW networks
\cite{WATTS98,WATTS99,NEWMAN00,NEWMAN99,MONASSON99,SCALETT91,BARRAT00,GITTERMAN00,KIM01,
PEKALSKI01,HONG02_1,HONG02_2,HERRERO02,JEONG03,Toral,HASTINGS03,BLUMEN_2000a,BLUMEN_2000b,
ALMAAS_2002,KOZMA03,KOZMA05,KOZMA05b,HASTINGS04}. Of these, the
ones most closely related to our work are the XY model
\cite{KIM01}, the EW model \cite{KOZMA03,KOZMA05,KOZMA05b},
diffusion
\cite{MONASSON99,BLUMEN_2000a,BLUMEN_2000b,ALMAAS_2002,KOZMA03,KOZMA05,KOZMA05b,HASTINGS04},
and current flow \cite{KORN_PLA_swrn} on SW networks. The findings
suggest that systems without inherent frustration exhibit (strict
or anomalous)
\cite{KOZMA03,KOZMA05,KOZMA05b,HASTINGS03,HASTINGS04}
mean-field-like behavior when the original short-range interaction
topology is modified to a SW network. In essence, the SW
couplings, although sparse, induce an effective relaxation to the
mean of the respective local field variables, and in turn, the
system exhibits a mean-field-like behavior \cite{HASTINGS03}. This
effect is qualitatively similar to those observed in models with
``annealed'' long-range random couplings
\cite{DRT90,BR91,BLUMEN_2000b}, but on (quenched) SW networks, the
scaling properties can differ from those with annealed
interactions
\cite{KOZMA03,KOZMA05,KOZMA05b,HASTINGS03,HASTINGS04}.

\end{section}

\begin{acknowledgments}
We acknowledge the financial support of NSF through DMR-0113049, DMR-0426488, and the
Research Corporation through RI0761. Z.T. was supported by DOE through W-7405-ENG-36.
Z.R. has been supported in part by the Hungarian Academy of Sciences through Grant OTKA-T043734.
\end{acknowledgments}

\end{document}